\documentclass[]{aa}
\usepackage{graphicx}
\usepackage[bookmarks=false]{hyperref}
\usepackage[varg]{txfonts}
\usepackage{natbib}
\usepackage{siunitx}
\usepackage{multirow}
\usepackage{color}
\usepackage{url}
\usepackage{mathtools}
\usepackage{float}
\usepackage{booktabs} % For \toprule, \midrule and \bottomrule
\usepackage{pgfplotstable} % Generates table from .csv
\pgfplotsset{compat=1.15}
\makeatletter
\renewcommand*\aa@pageof{, page \thepage{} of \pageref*{LastPage}}
\makeatother

\newcommand{\figref}[1]{Figure~\ref{#1}}

\newcommand{\lya}{Ly-$\alpha$ }

\newcommand{\lo}{\lambda_{\rm{opt}} }

\newcommand{\fnew}{q_{4000}(\lambda)}
\newcommand{\fold}{q_{5400}(\lambda)}
\newcommand{\be}{\begin{equation}}
\newcommand{\ee}{\end{equation}}

\newcommand{\secref}[1]{Section~\ref{#1}}
\newcommand{\mgt}{\textcolor{magenta}}
\newcommand{\tabref}[1]{Table~\ref{#1}}

\begin{document}

\title{Spectral shape corrections for SDSS BOSS quasars}

\author{Dinko Milakovi{\'c}\inst{1}
  \and John K. Webb\inst{2}
  \and Chung-Chi Lee\inst{2}
  \and Evgeny O. Zavarygin\inst{3}}

\institute{INAF, Osservatorio Astronomico di Trieste, Via G.\,B. Tiepolo 11, I-34143 Trieste, Italy 
\and Clare Hall, University of Cambridge, Herschel Road, Cambridge CB3 9AL, UK
\and AdServer Ltd, 37 Professora Popova Street, Saint-Petersburg, 197022, Russia.
}

\date{Received \phantom{mmmmmmm} / Accepted \phantom{mmmmmmm}}

\abstract{Modifications were made to the Sloan Digital Sky Survey's Baryonic Oscillations Spectroscopic Survey (SDSS/BOSS) optical fibres assigned to quasar targets in order to improve the signal-to-noise ratio in the \lya forest. However, the penalty for these modifications is that quasars observed in this way require additional flux correction procedures in order to recover the correct spectral shapes. In this paper we describe such a procedure, based on the geometry of the problem and other observational parameters. Applying several correction methods to four SDSS quasars with multiple observations permits a detailed check on the relative performances of the different flux correction procedures. We contrast our method (which takes into account a wavelength dependent seeing profile) with the BOSS pipeline approach (which does not). Our results provide independent confirmation that the geometric approach employed in the SDSS pipeline works well, although with room for improvement. By separating the contributions from four effects we are able to quantify their relative importance. Most importantly, we demonstrate that wavelength dependence has a significant impact on the derived spectral shapes and thus should not be ignored.
}

\maketitle 

\section{Introduction}
\label{sec:intro}

Absorption by neutral hydrogen leaves a series of absorption lines imprinted on the spectra of distant quasars, known as the \lya forest \cite[e.g.][]{Lynds1971,Sargent1980,Weymann1981,Rauch1998}. This gas traces the matter distribution on Gpc scales and can be used to study the distribution and evolution of baryons over cosmological timescales \citep[e.g.][]{Kim2002,Penton2004,Lehner2007,Zavarygin2019}. Large statistical samples of the \lya forest are therefore important tools for cosmological studies. The Baryonic Oscillations Spectroscopic Survey \citep[BOSS,][]{Ahn2012}, a part of the Sloan Digital Sky Survey \citep[SDSS,][]{York2000}, provides the largest available sample of medium-resolution quasar spectra.

BOSS quasar spectra \citep{Dawson2013}, are known to suffer from at least two systematic effects, the ``Balmer problem'' and the ``fibre offset problem'' \citep{Lee2013}. The Balmer problem arises from stellar absorption not being properly accounted for when flux calibrating the quasar spectra, such that the derived flux calibration correction function has residual stellar absorption features which can be transferred to the quasar spectra. The fibre offset problem for BOSS quasar spectra exists because quasars and photometric calibrating stars were fibre-centered at different central wavelengths. This was done in order to maximise the signal-to-noise ratio (S/N) in the \lya forest. However, neglecting to correct for it results in incorrect quasar spectral shapes \citep{Lee2013}. For this reason, a geometric procedure was implemented \citep{Margala2016} to correct quasar spectral shapes.

In fact, prior to seeing the comprehensive analysis of \citet{Margala2016}, we had already developed a similar (but slightly simpler) method which was being prepared for publication. Once the Margala procedure had been published, we saw little point in publishing our own, and shelved the work. However, during a more recent detailed analysis of \lya~forest data to constrain cosmological isotropy \citep{Zavarygin2019}, we discovered a systematic in SDSS BOSS spectra that emulates anisotropy and appears to be, at least in part, generated by the application of the fibre offset correction method of \cite{Margala2016}. This motivated us to re-examine the issue, hence the present paper in which we present a new, independent geometric correction method. The method we introduce here and that of Margala adopt the same basic geometric approach. However, the Margala method includes corrections to account for a quasar's plate position, whereas ours does not (since some of the relevant technical data was not available to us). On the other hand the Margala method assumes a constant seeing profile whereas ours does not and instead allows for wavelength dependence.

The rest of this paper is structured as follows: in \secref{sec:problem} we describe the geometry of the problem, the modifications made to the BOSS optical fibres assigned to quasars, and the mathematical model used to correct the fibre offset problem. We show that the effect of wavelength dependant seeing is significant and must be taken into account.
In \secref{sec:sampleselection} we define selection procedures to identify quasars with multiple exposures since these enable a direct test of how well the correction procedures work. The stringent and specific selection results in only four suitable quasars in the SDSS samples. The results of applying various correction procedures are described in \secref{sec:application}. \secref{sec:discussion} discusses the findings of this study and the Appendices provide corresponding details for three quasars, only one quasar being discussed in the main text.

\section{Fibre offset geometry and corrections} \label{sec:problem}
\subsection{The BOSS spectrograph}

BOSS is a component of the SDSS-III survey \citep{Eisenstein2011,Ahn2012}. SDSS BOSS is conducted on a 2.5m Sloan telescope \citep{Gunn2006} at Apache Point Observatory in New Mexico, providing the spectra of $\sim 1.5$ million luminous red galaxies up to $z=0.7$ and $\sim$ \SI{150000}{} quasar spectra at $z>2$ \citep{Dawson2013}. Quasar targets for the spectroscopic survey are selected based on the object colours, derived from earlier photometric imaging (see \citealt{Ross2012} for details on how quasars are selected for spectroscopy). The 13$^{\text{th}}$ BOSS data release (DR13) contains more than \SI{294000}{} medium resolution ($R=\lambda/\Delta\lambda \approx1500 - 2500$, where $\Delta\lambda$ is the full-width at half-maximum of the one-dimensional point spread function) quasar spectra, covering more than 10,400 $\rm{deg}^2$ \citep{Alam2015}.

The BOSS spectra are obtained using twin spectrographs covering a wavelength range $\sim 3000 - \SI{11000}{\angstrom}$. The spectrographs are fed by 1000 optical fibres (500 per spectrograph), each subtending \SI{2}{\arcsecond} on the sky. Further details about the spectrograph design can be found in \cite{Smee2013}. Observations were designed to maximise the fraction of BOSS scientific targets assigned a fibre. A maximum of 900 fibres were assigned to scientific targets such as quasars. The remainder was reserved for calibration stars and sky background observations \citep{Dawson2013}.

\subsection{The fibre offset problem}\label{sec:problemdetails}

Atmospheric differential refraction will cause the centre of the quasar seeing profile vary with wavelength and altitude on the sky. In order to obtain more precise data on the \lya forest (i.e. to improve the S/N at wavelengths $\lambda \lesssim \SI{4000}{\angstrom}$  \citep{Dawson2013,Lee2013}, the mechanics of attaching the optical fibres used for quasar observations to BOSS plate positions was modified in two ways: (1) thin wafers of sizes $175 - 300 \; \mu\rm{m}$ (the size of the wafer depends on the distance of the fibre plug to the plate center) were attached to the plug holes to provide an offset to the focal plane along the $z$-axis (i.e. towards the target); (2) the positions of the fibres were also offset by up to $\sim\ang{;;0.5}$ in the focal plane (i.e. along the $x$- and $y$-axes) in order to maximise the light entering the fibre. These fibre adjustments were thus both lateral and axial and designed such that the target image and fibre centre coincide at \SI{4000}{\angstrom} for quasars and \SI{5400}{\angstrom} for calibration stars.

\begin{figure}[t]
\begin{center} 
	\includegraphics[width=0.8\columnwidth]{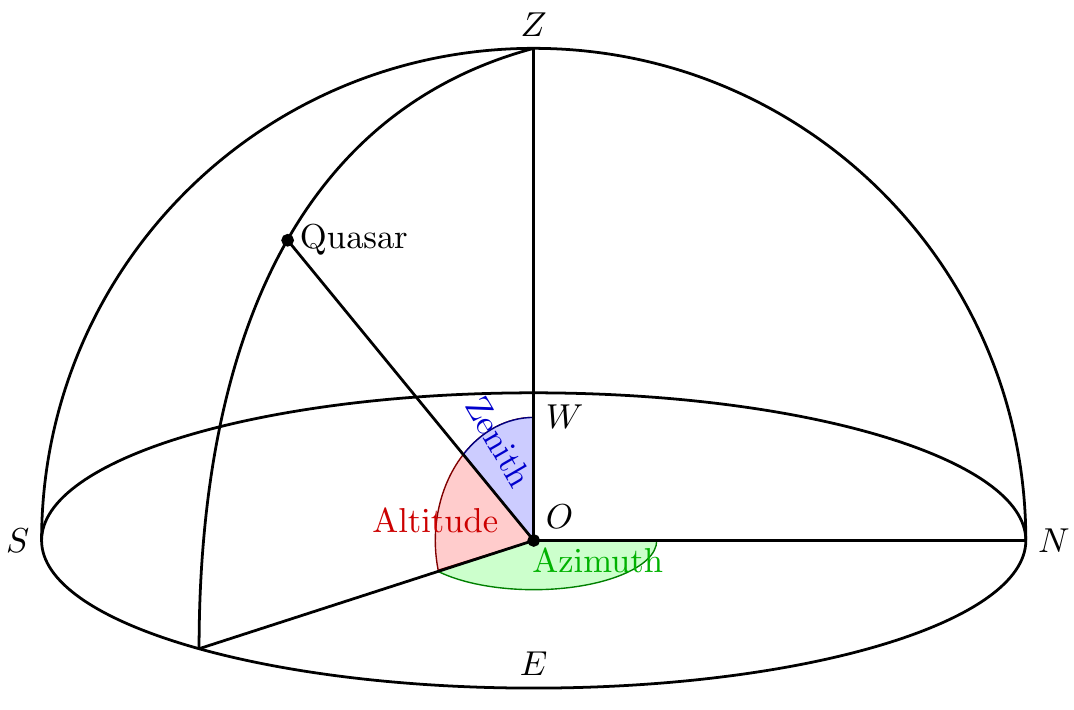}
	\caption{The coordinate system showing the positions of the observer ($O$) and the source (quasar) on the sky. The zenith angle, altitude, and azimuth are shown in blue red, and green, respectively.}
	\label{fig:zenith}
\end{center}
\end{figure}

Atmospheric differential refraction is a function of wavelength and of the air mass towards the observed object. The relative air mass is the total atmospheric path length towards the observed object in units of path length at the zenith. To first order, relative air mass is given by:
\be
X = \sec Z,
\ee
where $Z$ is the zenith angle. Atmospheric differential refraction will affect those objects with larger air mass more strongly that those with low air mass. Therefore, the fibre offset problem is more pronounced for quasars observed at large zenith angles, as is shown in \secref{sec:4sensitivities}. The positions of the observed object (quasar), the zenith, and the observer are illustrated in \figref{fig:zenith}.

\begin{figure}[t]
\begin{center} 
\includegraphics[trim={0 5cm 0 5cm}, clip, width = 0.8\columnwidth]{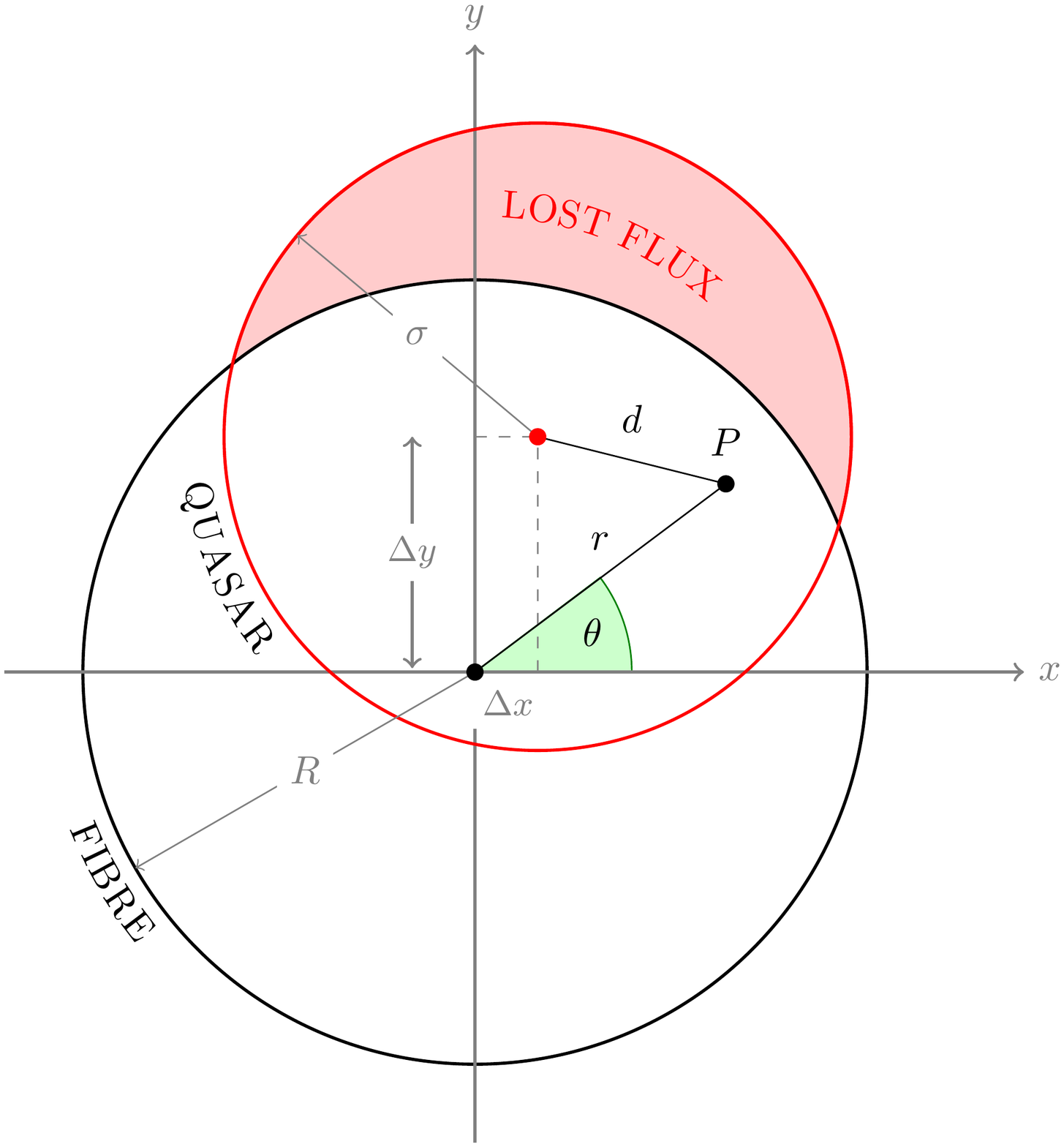}
\caption{The plane perpendicular to the sky direction, which points out of the page. The black circle represents the fibre with a radius $R$. The red dot and circle represent the quasar position and profile, i.e. the $\sigma$ of the quasar Gaussian seeing profile. The red shaded area illustrates the light lost. To calculate the captured flux, we integrate the normalised quasar light intensity within the fibre radius. The figure is adapted from \cite{Barnes1988}.}
\label{fig:fibre}
\end{center}
\end{figure}

\figref{fig:fibre} illustrates the optical fibre/quasar geometry. For ease of calculation, we position the origin of the coordinate system at the fibre centre at $\lambda=\lambda_0$, with axes defined as in \figref{fig:fibre}. We will refer to this simply as	the fibre centre. The quasar seeing profile has a radius $\sigma = \textrm{FWHM}/2\sqrt{2\ln 2} \approx \textrm{FWHM}/2.35$, where FWHM is the full-width at half-maximum intensity of the quasar seeing profile. Telescope guidance imperfections, perturbations in temperature, humidity and pressure of the atmosphere, as well as deviations from estimated observation times, will cause the quasar centre to move by $\Delta x$ along the $x$-axis. The quasar centre at $\lambda\ne\lo $ is offset by $\Delta y$ along the $y$-axis due to atmospheric differential refraction \citep{Filippenko1982}. The quasar profile will therefore lie partially outside of the fibre area for large enough values of $\Delta r = \sqrt{\Delta x^2 + \Delta y^2}$. From this, it follows that a fraction of the incoming quasar light will be lost at certain wavelengths and the flux underestimated \citep{Barnes1988}. \par

\begin{figure*}[t]
\begin{center} 
\includegraphics[width=\linewidth]{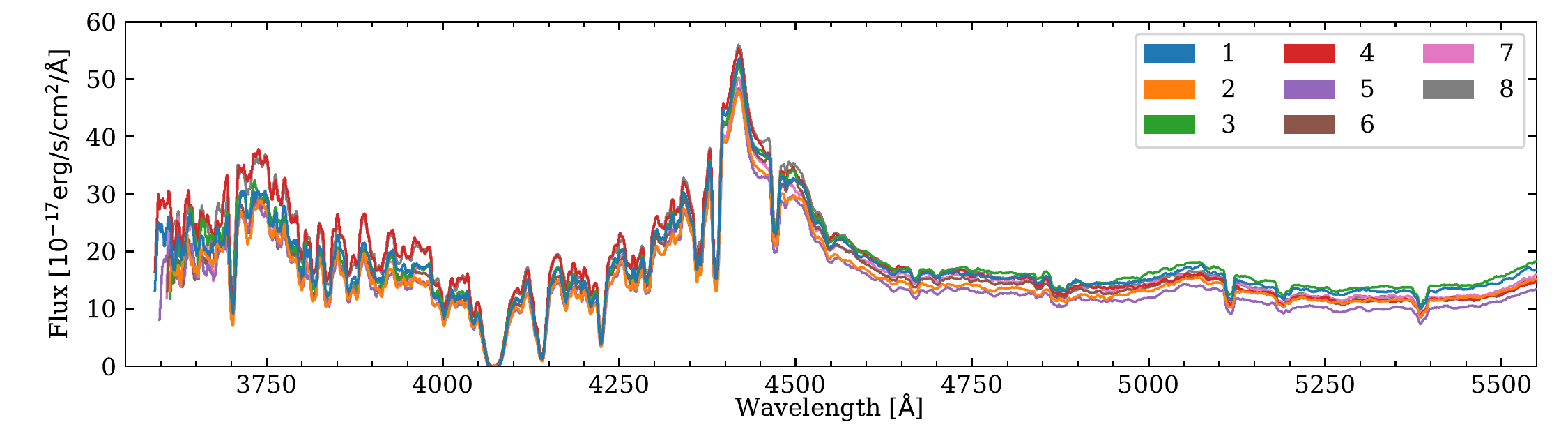}
\caption{Eight uncorrected SDSS BOSS observations of the quasar J022836.08+000939.2 (Thing ID = 97003678) in DR13, smoothed by a box-car filter of 10 pixel width. The quasar was observed between September 2011 and October 2013. The spectral shape of the quasar continuum is seen to vary for different observations. If spectral index and/or absolute flux do not change,  this variation probably occurs due to a fibre positioning offset, as first noted by \citet{Lee2013} and discussed in this paper in \secref{sec:problem}. The numbers in the legend refer to the observations listed in \tabref{tab:qso_exposures}.}
\label{fig:bad_spectrum}
\end{center}
\end{figure*}

Figure \ref{fig:bad_spectrum} shows eight BOSS exposures of the quasar J022836.08+000939.2. The \cite{Margala2016} correction has not been applied. The spectral shapes are seen to differ for individual observations. Whilst cannot be sure that quasar variability is not in part responsible for the observed spectral shape differences, the fibre offset problem will also contribute to the apparent spectral shape variations.

\subsection{Calculating the fibre offset $\Delta y$}

We now consider a general case in which the fibre is centred at wavelength $\lambda_0$, and calculate the flux coming from the quasar which is collected by the fibre as a function of wavelength. This requires knowledge of the distance of the two centres (the target and the fibre) along the $y$-axis, denoted by $\Delta y$ (see \figref{fig:fibre}). $\Delta y$ is determined by the zenith angle, $Z$, and the difference in the refractive indices of air, $n$, at wavelengths $\lambda$ and $\lambda_0$. $\Delta y$ (measured in arcseconds), is then calculated as follows \citep{Smart1931,Filippenko1982}:
\begin{align}
\Delta y (\lambda,\lambda_0)& =  y(\lambda) - y(\lambda_0) \nonumber \\ & \approx 206265\left[ n(\lambda) - n(\lambda_0)\right] \tan Z,
\label{eq:Dy}
\end{align}
The refractive index, obtained using lunar laser ranging, \citet{Marini1973}, is given by 

\be
\begin{split}
n (\lambda, P, e, T) = \left(287.604 + \frac{1.6288}{\lambda^2} + \frac{0.0136}{\lambda^4} \right) \times \\
\times \left(\frac{P}{1013.25}\right) \left(\frac{1}{1 + 0.003661 T}\right) \\
- 0.055\left(\frac{760}{1013.25}\right) \left(\frac{e}{1 + 0.00366 T}\right)
\end{split}
\label{eq:n}
\ee
where,
\begin{tabbing}
\= $\lambda$ \= = \= wavelength in microns, \\
\> $P$ \> = \> atmospheric pressure in millibars,\\
\> $e$ \> = \> partial water vapour pressure in millibars,\\
\> $T$ \> = \> air temperature in degrees Celsius.\\
\end{tabbing}
The SDSS FITS header for each quasar exposure provides the atmospheric pressure in inches of mercury. Equation \ref{eq:n} requires $P$ in millibars, given by $P(\textrm{mb}) = 33.86389\, P(\textrm{inches of mercury})$. 

The second term (involving $e$) in Equation\eqref{eq:n} cancels in the subtraction of Equation\eqref{eq:Dy} so in practice is not needed, but for completeness we illustrate the full calculation of $n(\lambda, P, e, T)$. The SDSS FITS headers provide the relative humidity, H, for each exposure. The partial water vapour pressure is
\begin{equation}
e = {\textrm H} e'
\end{equation}
where
\be
\begin{split}
e'(P,T) = &\left[1.0007 + (3.46 \times 10^{-6} P)\right] \\
& \times 6.1121\exp\left[\dfrac{17.502 \; T }{240.97 + T} \right].
\end{split}
\label{eq:refractiveindex}
\ee
\citep{Buck1981}.

We evaluate $\Delta y$ using Equation \eqref{eq:Dy} for each spectrum over the observed wavelength range.

\subsection{Flux correction calculation} \label{sec:fcc}

Equation \eqref{eq:Dy} provides the physical offset of the quasar seeing profile from the fibre centre as a function of wavelength. The SDSS pipeline data reduction procedure assumes (incorrectly) that the quasar and calibration stars are both centred at $\lambda_0=\SI{5400}{\angstrom}$. The flux correction function derived here corrects for this assumption, as does the correction function of \cite{Margala2016}.

However, here we take into account an additional point that is not included in the SDSS pipeline correction; the fibre offset corrections derived in \cite{Margala2016} assume that the seeing profile (measured empirically for each exposure at \SI{5000}{\angstrom} is constant with wavelength. The correction function described in this paper allows for wavelength dependent according to 
\be
\label{eq:seeing}
\sigma = \sigma_0 \left( \frac{\lambda}{{\SI{5000}{\angstrom}}}\right)^{-0.2} 
\ee
e.g. \cite{Roddier1981,Meyers2015}, where $\sigma_0$ is the 
seeing measured by the SDSS secondary camera during observations using a filter with maximum throughput at \SI{5000}{\angstrom}. 

The normalised quasar intensity is, in Cartesian coordinates,
\be
\rho(x,y) = \frac{1}{2\pi\sigma^2}e^{-\left[(x - \Delta x)^2+(y - \Delta y)^2\right]/2\sigma^2},
\ee
where $x$, $\Delta x$, $y$, $\Delta y$, are illustrated in Figure \ref{fig:fibre}. 

Transforming to polar coordinates, the normalised quasar intensity becomes:
\be
\rho(r,\theta) = \frac{1}{2\pi\sigma^2}e^{-d^2(r,\theta)/2\sigma^2},
\ee
where $d$ is the distance from the quasar centre and is a function of $r$ and $\theta$ (illustrated in Figure \ref{fig:fibre}). 

Atmospheric refraction shifts the quasar centre in the $y$ direction only. Any fibre-quasar misalignments perpendicular to the atmospheric refraction direction, $\Delta x$, can also in principle create spectral shape changes, since the seeing profile depends on wavelength. However, we assume that $\Delta x$ is small compared to refraction offsets from Equation \eqref{eq:Dy} thus assume $\Delta x =0$.

The law of cosines relates $d$ to the fibre offset $\Delta y$ and the coordinates of point $P$, $r$ and $\theta$,
\be
d^2 = r^2 + \Delta y^2 - 2r\Delta y \sin{\theta}.
\label{eq:R}
\ee
Allowing for the fibre truncating the Gaussian (asymmetrically when $\lambda \ne \lambda_0 $), the fraction of incident quasar light entering the fibre is
\be
I = \int_0^{2\pi} \int_{0}^{R} \rho(r,\theta) r dr d\theta = \int_0^{2\pi} \int_{0}^{R} \frac{1}{2\pi\sigma^2}e^{-d^2(r,\theta)/2\sigma^2} r dr d\theta,
\label{eq:ii}
\ee
where $R$ is the fibre radius. Expressed as a function of $\lambda$ and $\lambda_0$, Equation \eqref{eq:ii} is
\be
I(\lambda,\lambda_0) = \int_0^{2\pi} \int_{0}^{R} \frac{1}{2\pi\sigma^2}e^{-d^2(\lambda,\lambda_0,r,\theta)/2\sigma^2} r dr d\theta,
\label{eq:iii}
\ee

We now define the throughput functions as
\be
\fold = \dfrac{I(\lambda, \SI{5400}{\angstrom})}{I(\SI{5400}{\angstrom}, \SI{5400}{\angstrom})}
\ee
and
\be
\fnew = \dfrac{I(\lambda, \SI{4000}{\angstrom})}{I(\SI{4000}{\angstrom}, \SI{4000}{\angstrom})},
\ee
which we calculate for each quasar. $I(\SI{5400}{\angstrom}, \SI{5400}{\angstrom})$ and $I(\SI{4000}{\angstrom}, \SI{4000}{\angstrom})$ signify the quasar light collected at $\lambda = \lambda_0$, i.e. when quasar's centre coincides with the fibre's centre, and serve as normalisation constants.

The correction function, $C$, is the ratio of the throughput functions,
\be \label{eq:Clambda}
C(\lambda) = \dfrac{\fold}{\fnew}.
\ee \par
The corrected quasar flux $f(\lambda)$ then becomes
\be\label{eq:proper_flux}
f(\lambda) = C (\lambda) \, f'(\lambda),
\ee
where $f'(\lambda$) is the uncorrected quasar flux at wavelength $\lambda$.

\subsection{Sensitivity of the correction function to different variables} \label{sec:4sensitivities}

Section \ref{sec:fcc} illustrates that the fibre offset correction calculation depends on air pressure, temperature, seeing, and zenith angle. It is therefore of interest to examine the relative importance of these parameters. To this end we calculate Equation \eqref{eq:Clambda} for representative ranges in each parameter, keeping 3 parameters constant and varying only one at a time. The parameter ranges used for this purpose were obtained empirically from the observed ranges seen in the quasar sample described shortly. 

\begin{figure*}
    \centering
    \includegraphics[width=\textwidth]{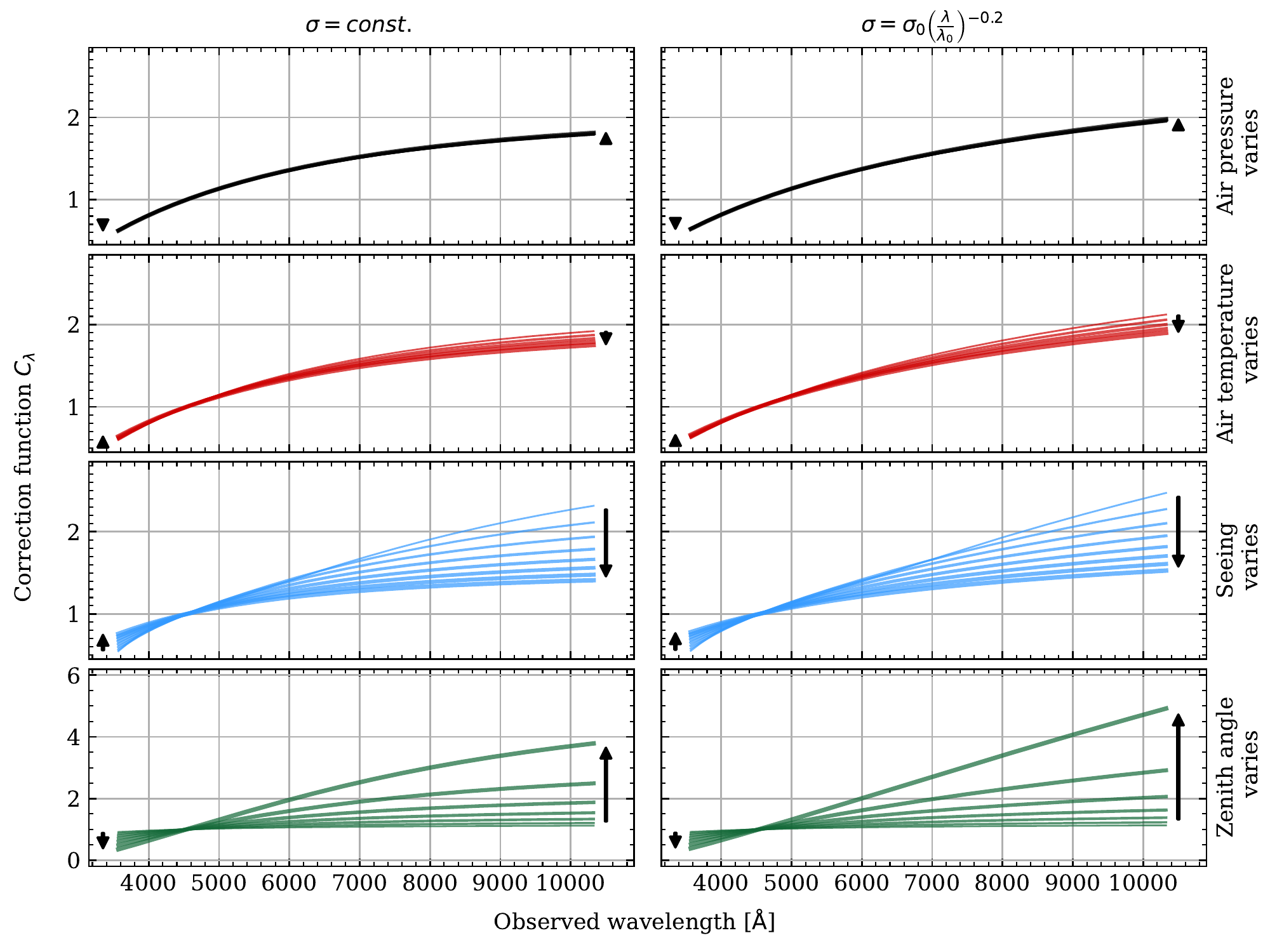}
    \caption{These plots illustrate the relative importance of the variables air pressure, air temperature, seeing, and zenith angle. In the left column, seeing is independent of wavelength whilst the right column shows the results for wavelength dependent seeing. The varying parameter is indicated on the right hand side. The other three parameters in each panel are set at mean values using non-repeated MJD settings in Table \ref{tab:qso_exposures}. The arrows in each panel indicate the direction in which the correction function moves as the parameter value increases. Line thickness also increases with the increase in the parameter value. See Section \ref{sec:4sensitivities} for details.}
    \label{fig:4params}
\end{figure*}

\figref{fig:4params} illustrates the results: from top to bottom: air pressure ($P$, black curves), air temperature ($T$, red curves), atmospheric seeing ($\sigma$, blue curves), and zenith angle ($Z$, green curves). In each panel, only one parameter is allowed to vary and the non-varying parameters are held fixed. The left hand column illustrates $C(\lambda)$ for constant seeing whilst the right hand column shows the results for wavelength dependent seeing as per Equation \ref{eq:seeing}. The non-varying parameters are fixed at their average values, taken over the sample of observations used in this work: $p=\SI{729.91}{\milli\bar}$, $T=\SI{8.58}{\celsius}$, $\sigma=\SI{1.58}{\arcsecond}$, $Z=\SI{39.11}{\degree}$. 
The following values are used for (a) pressure: 726, 728, 730, and \SI{734}{\milli\bar}; (b) temperature: -5, 0, 5, 10, and \SI{15}{\celsius}; (c) seeing: 1.0, 1.2, 1.4, 1.6, 1.8, 2.0, 2.2, and \SI{2.4}{\arcsecond}; and (d): zenith angle: 20, 25, 30, 35, 40, 45, and \SI{50}{\degree}. Note the different $y$-axis scale for the bottom panel, when zenith is varied. The correction function evidently depends most sensitively on the zenith angle, with seeing being the next most sensitive parameter. Variations in temperature and pressure have the least effect.

\section{Selecting SDSS quasars with multiple exposures} \label{sec:sampleselection}
\begin{table*}
    \centering
    \caption{Details on the quasars used in this work. The first three columns give the quasar name, its unique identifier in the SDSS DR13 (``Thing ID''), and in the SDSS photometric database (``Obj ID''). The following three columns give the quasar's right ascension and declination (in degrees) and the quasar's emission redshift. The final column gives the number of observations which satisfy the criteria described in Section \ref{sec:sampleselection}.}
    \label{tab:quasar_details}
    \begin{tabular}{lrrrrcc}
    
    \toprule\toprule
         \multicolumn{1}{c}{SDSS Name} & \multicolumn{1}{c}{Thing ID} & \multicolumn{1}{c}{Obj ID} & {RA (deg)} & {Dec (deg)} & $z_{em}$ & $N_{obs}$ \\
         \midrule
         J022954.42$-$005622.5 & 74944092 & 1237666406848200831 & $37.48$ & $-0.94$ & 2.3107 & 6\\
         J023308.31$-$002605.0 & 82655414 & 1237657070090125331 & $38.28$ & $-0.44$ & 2.4965 & 7 \\
         J022836.08$+$000939.2 & 97003678 & 1237663784213676145 & $37.15$ & $0.16$ & 2.6324 & 8 \\
         J023259.60$+$004801.7 & 110619413 & 1237657587096813739 & $38.25$ & $0.80$ & 2.2272 & 6\\
         \toprule\toprule
    \end{tabular}

\end{table*}

\begin{figure*}[t]
\begin{center}
\includegraphics[width=1.02\textwidth]
{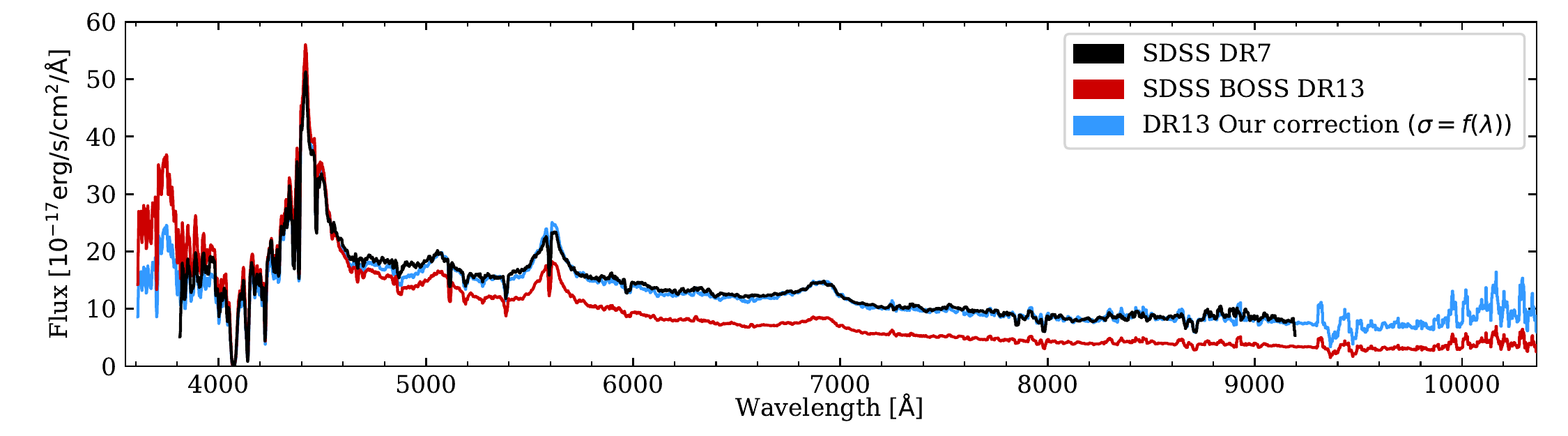}
\caption{Corrections of a single observation. The BOSS spectrum of J022836.08+000939.2 (Plate: 3647, MJD: 56596, Fibre: 776) before (red) and after (blue) correcting for the fibre offset problem. The SDSS-I observation (Plate: 2635, MJD: 54114, Fibre: 621), unaffected by the fibre offset problem, is shown in black.}
\label{fig:spec_corr}
\end{center}
\end{figure*}

In order to examine the impact of applying Equation \eqref{eq:Clambda}, we use a sample of SDSS DR13 quasars. More specifically, we choose quasars having multiple  exposures with good signal-to-noise ratios (S/N). By applying the correction to each exposure and then examining internal consistency for each set for a given quasar, we may be able to quantify any potential spectral shape improvement resulting from our fibre offset correction. The same sample can also be used to check for consistency between the \cite{Margala2016} and our corrections. We also compare results obtained using corresponding DR16 spectra \citep{Ahumada2020} which have had the Margala correction applied but to individual 15 minute exposures (rather than only to summed one hour spectra).

The quasar selection criteria used for this purpose are contained within the file \texttt{ spAll-v5\_9\_0.fits}\footnote{\url{https://data.sdss.org/sas/dr13/eboss/spectro/redux/v5_9_0/spAll-v5_9_0.fits}} (a summary file of metadata such as target coordinates, redshifts, photometry, identifiers, classification, number of exposures, etc.). The eight selection criteria applied are:
\begin{enumerate}
\item LAMBDA\_EFF = 4000 (a modified fibre, centred at $\lambda=\SI{4000}{\angstrom}$, was used)
\item OBJTYPE = "QSO" (the object was targeted as a quasar)
\item CLASS = "QSO" (the object was subsequently classified as a quasar)
\item ZWARNING = 0 (no known issues with redshift estimate)
\item NSPECOBS $>4$ (the quasar was observed more than four times)
\item SN\_MEDIAN\_ALL $>$ 10 (S/N per observation $>$ 10 over the entire wavelength range and for all observations)
\item ANCILLARY\_TARGET1 = 0 (not targeted as a part of any quasar variability or broad absorption line programmes)
\item MJD $>$ 55447 (the date after which weather information was routinely saved into FITS file headers)
\end{enumerate}

These stringent criteria result in a sample of 9 quasars. For five of these quasars, we could not get a good power-law fit to the quasar continuum regions (Section \ref{sec:si}). The spectrum of one resembled a type 1 AGN (``Thing ID'' 97024300) and one was a strong broad absorption line quasar (110633310). In the remaining three cases (70363665, 78014536, and 106364494), a power law gave a reasonable fit to the uncorrected BOSS spectra but after applying the correction functions, a power law no longer produced a good fit, for reasons unknown. This left us with only four quasars, which we now discuss. Basic information on the four quasars is given in \tabref{tab:quasar_details}, together with their unique identifiers in the SDSS database, ``Thing ID'' and ``Obj ID''. We find that the Thing ID for the same quasar differs between DR12 and DR13, so we also included the Obj ID for each quasar. ObjID identifies an object in the SDSS image catalog used by the Catalog Archive Server\footnote{see \url{https://www.sdss.org/dr13/help/glossary/}}.

\section{Applying fibre offset corrections to 4 quasars with multiple BOSS exposures} \label{sec:application}

Ideally, we would be able to draw upon completely independent (and more precise) observations of the 4 quasars above, enabling comparisons with BOSS exposures before and after applying fibre offset corrections. Unfortunately, the available data permits this only in a limited sense. We thus proceed by exploring two quantities: (1) does applying fibre offset corrections produce greater consistency between the individual quasar exposures and (2) does it also result in more consistent spectral index estimates. Individual exposures of the four quasars we use are tabulated in \tabref{tab:qso_exposures}.

Assuming atmospheric differential refraction is the largest factor contributing to incorrect spectral shapes, we would expect that applying the flux correction procedure would result in consistent spectral shapes for individual exposures on the same quasar. 

\figref{fig:spec_corr} shows the BOSS observation of J022836.08+000939.2 (plate 3647, MJD 56596, fibre 776, 4$\times$15 minute exposures, co-added), before and after correcting for the fibre offset problem. The SDSS-I observation is also shown (unaffected by the fibre offset problem, plate 2635, MJD 54114, fibre 621). It can be seen that the corrected BOSS spectrum is in very good agreement with the SDSS-I spectrum. We comment shortly on the similarity between the flux correction method described in this paper and that of \citet{Margala2016}. On this basis of the result illustrated in \figref{fig:spec_corr}, one may infer that the correction works well. However, the interpretation is not straightforward as the following test shows.

\begin{figure*}[t]
\begin{center}
\includegraphics[width = 1.02\linewidth]{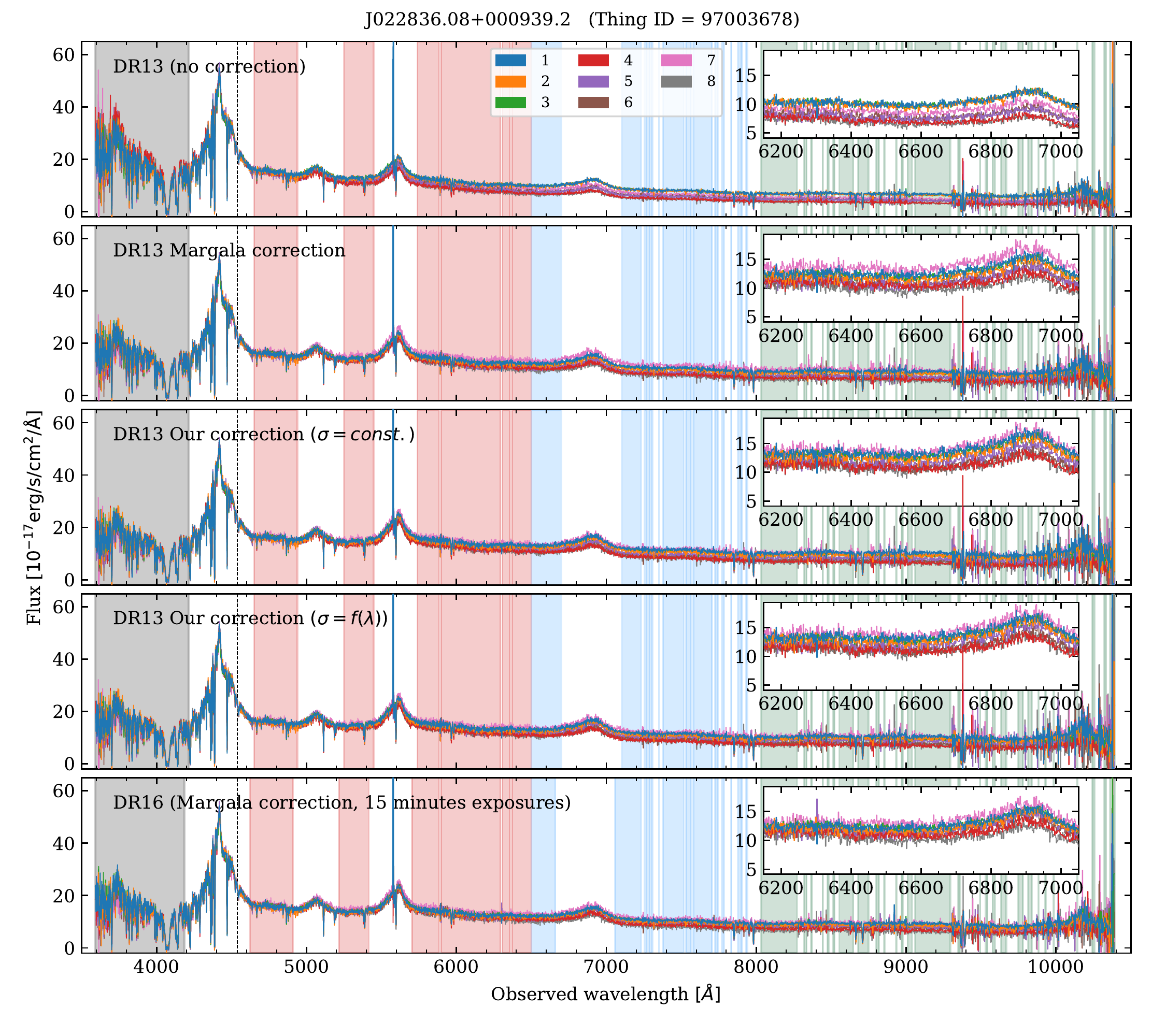}
\caption{Eight BOSS observations of the quasar J022836.08+000939.2 (Thing ID 97003678). Individual observations are shown as coloured lines, with the legend showing the observation ID in \tabref{tab:qso_exposures}. The wording within each box shows which correction has been applied (no correction has been applied in the top box, DR13). The inset in each panel illustrates a shorter wavelength region in more detail, giving a clearer view of the scatter between multiple exposures. The shaded areas show the regions used to calculate $\xi^2$ and its associated error, Equations \eqref{eq:xisquared} and \eqref{eq:sigxisq}, results listed in Table~\ref{tab:consistency_97176486}. The different colours correspond to the four wavebands used. The vertical dashed line around \SI{4538}{\angstrom} shows the point about which spectra are normalised to a common flux. The spike at \SI{5600}{\angstrom} is due to a cosmic ray hit.
}
\label{fig:spec_obs}
\end{center}
\end{figure*}

\begin{figure*}[htbp]
\begin{center}
\includegraphics[width = \linewidth]{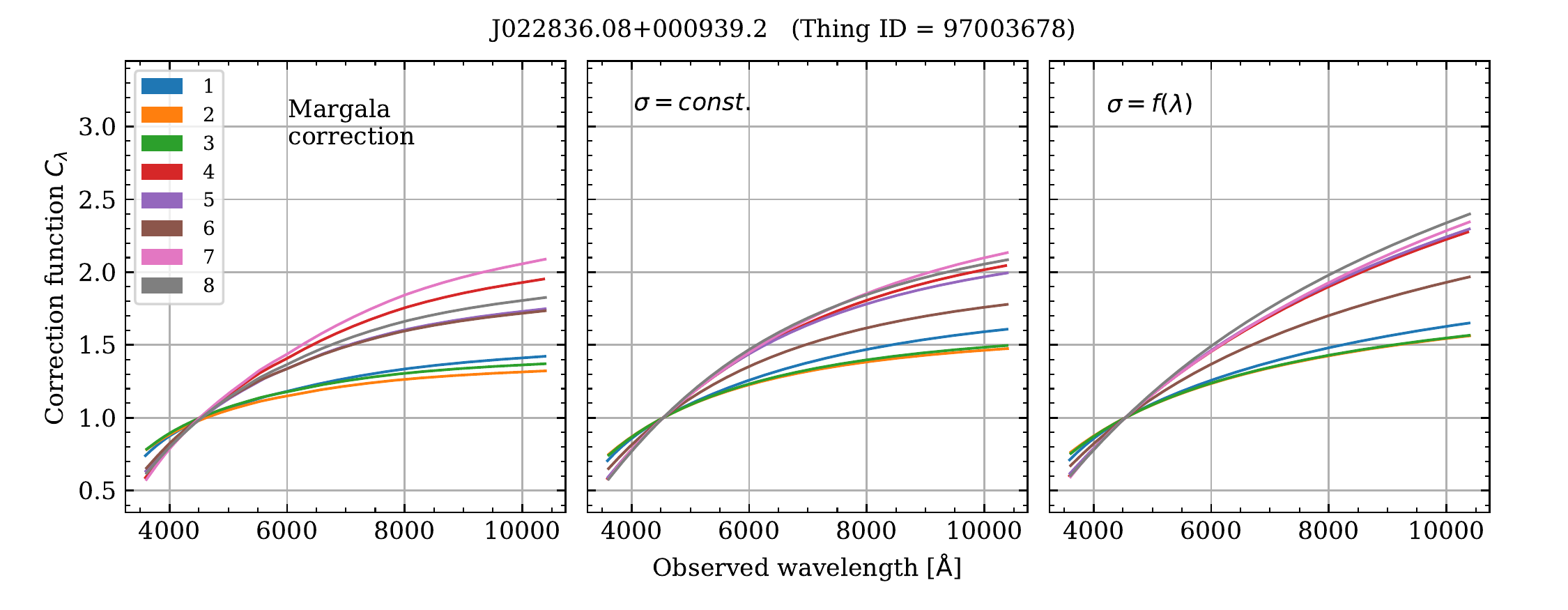}
\caption{Correction functions (Equation \eqref{eq:Clambda}) for the eight observations of J022836.08+000939.2 shown in \figref{fig:spec_obs}. The numbers in the legend correspond to the observation IDs in \tabref{tab:qso_exposures}.}
\label{fig:corr}
\end{center}
\end{figure*}

\figref{fig:spec_obs} illustrates all eight BOSS observations of J022836.08+000939.2, before and after applying our correction for the fibre offset problem. The correction functions for this set of observations are shown in \figref{fig:corr}, which shows clearly that the effect of wavelength dependence is significant. The effect the corrections have on the spectral shapes of the observations is evident from \figref{fig:corr}; the corrections lower the fluxes at wavelengths $\lambda \lesssim \SI{4538}{\angstrom}$, while increasing the fluxes at wavelengths $\lambda \gtrsim \SI{4538}{\angstrom}$. 

%https://ctan.ijs.si/tex-archive/graphics/pgf/contrib/pgfplots/doc/pgfplotstable.pdf
\begin{table*}
  \begin{center}
    \caption{List of quasar exposures used. The first column is a number to cross-reference to the curves shown in Figure \ref{fig:corr}, Tables \ref{tab:si_97176486}, \ref{tab:consistency_97176486}, and the figures and tables in the Appendices. The second column is an SDSS unique identifier (the three hyphenated parts are plate number-MJD-fibre). The remaining four columns are pressure in millibars, temperature in degrees Celsius, seeing in arcseconds, and zenith distance in degrees. Each group of exposures is preceded by the quasar SDSS Thing ID (see \tabref{tab:quasar_details}). }
    \label{tab:qso_exposures}
    \begin{filecontents*}{tables/csv/exposures.csv}
0.00,74944092,0.00,0.00,0.00,0.00
1,3615-55856-0214,731.54,10.00,1.21,33.87
2,3615-56219-0208,729.36,11.26,1.69,34.80
3,3615-56544-0220,728.64,11.19,1.51,33.86
4,3647-55827-0178,733.48,10.79,1.52,41.21
5,3647-55945-0132,727.95,2.44,1.94,44.08
6,3647-56568-0174,727.33,10.67,1.33,40.56
0.00,82655414,0.00,0.00,0.00,0.00
1,3615-55856-0008,731.54,10.00,1.21,33.87
2,3615-56219-0014,729.36,11.26,1.69,34.80
3,3615-56544-0002,728.64,11.19,1.51,33.86
4,3647-55827-0012,733.48,10.79,1.52,41.21
5,3647-56219-0010,729.17,9.45,1.76,40.51
6,3647-56568-0014,727.33,10.67,1.33,40.56
7,3647-56596-0014,726.08,2.83,1.81,43.74
0.00,97003678,0.00,0.00,0.00,0.00
1,3615-55856-0764,731.54,10.00,1.21,33.87
2,3615-56219-0774,729.36,11.26,1.69,34.80
3,3615-56544-0788,728.64,11.19,1.51,33.86
4,3647-55827-0772,733.48,10.79,1.52,41.21
5,3647-55945-0734,727.95,2.44,1.94,44.08
6,3647-56219-0768,729.17,9.45,1.76,40.51
7,3647-56568-0768,727.33,10.67,1.33,40.56
8,3647-56596-0776,726.08,2.83,1.81,43.74
0.00,110619413,0.00,0.00,0.00,0.00
1,3615-56219-0964,729.36,11.26,1.69,34.80
2,3615-56544-0952,728.64,11.19,1.51,33.86
3,3647-55827-0967,733.48,10.79,1.52,41.21
4,3647-55945-0999,727.95,2.44,1.94,44.08
5,3647-56219-0947,729.17,9.45,1.76,40.51
6,3647-56568-0941,727.33,10.67,1.33,40.56
\end{filecontents*}
\pgfplotstabletypeset[
      multicolumn names, % allows to have multicolumn names
      col sep=comma, % the seperator in our .csv file
      every head row/.style=
        {
        before row={\toprule  \toprule
                    %ID & Observation & Air pressure & Air temperature & Seeing & Zenith angle\\
                    },
        after row = {\midrule}
        }, 
        every last row/.style={
        %before row = \midrule,
        after row=\bottomrule\bottomrule},
        display columns/0/.style=
          {
          column name = ID,
          string type
          },
        display columns/1/.style=
          {
          column name=Observation,
          column type=c,
          string type
          },
        display columns/2/.style=
          {
          column name=$P (\SI{}{\milli\bar})$,
          column type=c,
          dec sep align
          },
        display columns/3/.style=
          {
          column name=$T (\SI{}{\celsius})$,
          column type=c,
          dec sep align
          },
        display columns/4/.style=
          {
          column name=$\sigma (\SI{}{\arcsecond})$,
          column type=c,
          dec sep align
          },
        display columns/5/.style=
          {
          column name=$Z (\SI{}{\degree})$,
          column type=c,
          dec sep align
          },
        fixed,
        fixed zerofill,
        precision=2,
        string replace={0.00}{},
        string replace={0}{},
        column type=c,
        %every row 0 every column 1/.style={postproc cell content/.style={@cell content=\textbf{##1}}}
        %clear infinite,
        %dec sep align,
        %every nth row={8}{before row=\midrule}
    ]{tables/csv/exposures.csv} % filename/path to file
  \end{center}
\end{table*}

\subsection{Consistency before and after correction} \label{sec:consistency}

One way to quantify the correction procedure effectiveness is to compare the consistency between multiple exposures of the same quasar, before and after correcting for the fibre offset problem. To do this we calculate the variance weighted departures from the mean flux squared\footnote{We avoid the usual $\chi^2$ notation to distinguish between $\chi^2$ in \secref{sec:si}, where we use it as the goodness-of-fit estimator.} within wavelength regions,
\be
\xi^2 = \sum_i \sum_j \left( \frac{ f_i^j - \langle f \rangle_i }{\sigma^j_i} \right)^2,
\label{eq:xisquared}
\ee
where the subscript $i$ corresponds to pixels within one spectrum and the superscript $j$ refers to different spectra of the same quasar. The quantity $\langle f \rangle_i$ is the ordinary (unweighted) mean flux in the $i^{th}$ pixel, averaged over all $j$ spectra of the same object.

Before calculating Equation~\eqref{eq:xisquared}, we normalise the observations of individual quasars to a common flux level in order to remove possible different flux levels remaining in the reduced spectra. These could be caused, for example, by time variations in the quasar continuum or by sky transparency variations. We do this by calculating the mean flux across all exposures in the region where the correction function crosses unity, in the wavelength range $\SI{4538}{}\pm\SI{10}{\angstrom}$, and normalising to a common value. We use this specific range (rather than the entire spectrum) because in this range we expect the spectral fluxes to be correct.

Furthermore, we exclude wavelength regions contaminated with strong quasar and/or sky emission lines from calculating the spread as per Equation \eqref{eq:xisquared}. The quasar emission lines we exclude are the Lyman-$\alpha$, Si\,{\tiny{IV}} and O\,{\tiny{I]}}, C\,{\tiny{IV}}, and C\,{\tiny{III]}} lines, falling in the following (quasar rest frame) wavelength ranges: $\SI{1160}{}\leq \lambda \leq \SI{1280}{\angstrom}$, $\SI{1360}{}\leq \lambda \leq \SI{1445}{\angstrom}$, $\SI{1500}{}\leq \lambda \leq \SI{1580}{\angstrom}$, and $\SI{1845}{}\leq \lambda \leq \SI{1955}{\angstrom}$ (respectively). We search for strong sky lines in the sky emission line atlas by \cite{Hanuschik2003}, containing 2810 emission lines identified in high-resolution ($R\approx \SI{45000}{}$) observations made by Ultraviolet and Visual Echelle Spectrograph (UVES). We identify 169 lines (in the \cite{Hanuschik2003} sample) which have a measured ${\rm FWHM} > \SI{0.15}{\angstrom}$ and flux $>5\times 10^{-16}{\rm erg\,s^{-1}\,cm^{-2}}$\,\AA$^{-1}$. In practice this means that we generally exclude the pixel containing the sky feature plus two pixels either side of each feature, i.e. we exclude approximately 2$\times$FWHM for each feature. Inspecting the data, this seemed to be appropriate. The initial pipeline processing of the data attempts to remove cosmic rays although inevitably some are likely to remain. We did not make any additional attempt to detect and remove cosmic rays.

To assess how well the correction works, the scatter between independent exposures of the same quasar prior to and after applying the correction can be compared, using the uncertainty on $\xi^2$,
\be
\sigma(\xi^2) = \sqrt{2K},
\label{eq:sigxisq}
\ee
where $K$ is the number of degrees of freedom, i.e. the number of pixels falling into the wavelength region across all considered spectra. We next discuss the results for each quasar individually.\\

\noindent{\it J022836.08+000939.2, Table \ref{tab:consistency_97176486}:}\\
The table provides the numerical values from Equations \eqref{eq:xisquared} and \eqref{eq:sigxisq} for the quasar J022836.08+000939.2. Measurements are made in four wavelength bands, as shown in the table. Corrections are listed prior to a fibre offset correction and after applying the Margala correction, our correction with constant seeing, our correction with wavelength dependent seeing, and the DR16 correction. The DR16 correction is the Margala correction except applied to individual 15 minute exposures prior to co-adding exposures rather than to summed one hour spectra. The bottom row in Table \ref{tab:consistency_97176486} gives the values obtained using the all four regions simultaneously i.e. it illustrates the overall benefit (or otherwise) of each correction method.

Inspecting the values of $\xi^2$ in Table \ref{tab:consistency_97176486}, the Margala correction and our comparable correction ($\sigma=const.$) provide only a slight improvement for the lowest wavelength region but make dramatic improvements for the three higher wavelength regions. Our second correction ($\sigma=f(\lambda)$) improves the consistency further in all four wavelength regions, as expected, given the clear importance of the wavelength dependant seeing correction (\figref{fig:4params}). The DR16 method improves consistency in three out of four regions (see discussion in the next paragraph). Overall (bottom row in Table \ref{tab:consistency_97176486}), the two $\sigma=const.$ methods perform almost identically, and the $\sigma=f(\lambda)$ and DR16 methods improve further (in that order).

Interestingly, the DR16 correction works better than the wavelength dependant seeing correction in the four highest wavelength regions but somewhat worse the lowest wavelength region. The reason for DR16 providing a worse correction for the lowest wavelength region is unknown but may be related to the lack of wavelength dependent seeing in the DR16 correction and/or an inaccurate absolute seeing value in the spectral FITS header. We have not investigated this further. The generally better performance of the DR16 correction for this quasar (i.e.\ in four out of five regions) highlights the importance of applying these corrections to individual rather than summed spectra. Presumably, if the DR16 correction included wavelength dependant seeing, even better agreement between individual exposures would be seen.\\

\noindent{\it J022954.42-005622.5, Table \ref{tab:consistency_75018778}:}\\
The Margala correction and our $\sigma=const.$ correction again perform very similarly, both improving on the uncorrected data consistency. Curiously (and unlike J022836.08+000939.2), our $\sigma=f(\lambda)$ correction fails to improve on the two $\sigma=const.$ corrections (with no obvious explanation). Moreover, the DR16 correction makes little or no improvement in the lowest two wavelength regions, but clearly improves consistency in the two highest wavelength regions. Overall, the DR16 correction again works best.\\

\noindent{\it J023308.31-002605.0, Table \ref{tab:consistency_827763272}:}\\
Again, the Margala correction and our $\sigma=const.$ correction again perform very similarly, both significantly improving on the uncorrected case. Our $\sigma=f(\lambda)$ correction is marginally worse than the two $\sigma=const.$ methods for the lowest wavelength region but improves consistency for three higher wavelength regions. The ordering of worst to best correction is again (Margala + our $\sigma=const.$ method), our $\sigma=f(\lambda)$, and DR16.\\

\noindent{\it J023259.60+004801.7, Table \ref{tab:consistency_110814178}:}\\
The results from this quasar gives peculiar results. All four fibre offset correction methods make the agreement between individual spectra conspicuously worse. The individual spectra are most consistent when no correction at all is applied. This can be seen clearly in \figref{fig:J023259correctionsB}. The effect gets worse as wavelength increases. We have no explanation for this. The other three quasars do not show this effect and in general, applying the fibre offset corrections improve consistency between individual spectra.\\

\noindent{\it Overall picture:}\\
Quasar 4 (J023259.60+004801.7) completely fails to co-operate with the correction methods. If for some reason the data had already been corrected prior to data release, this may explain what we find. However we have no evidence for that and the explanation seems unlikely. With little alternative, we will therefore assume (with no justification) that this quasar is anomalous and will proceed to draw general conclusions based on the (more consistent) remaining three cases. 

Overall, the Margala and our $\sigma=const.$ methods perform comparably. Our independent geometric correction thus supports the earlier procedures described in \cite{Margala2016}. Adding a wavelength dependant improves consistency. The DR16 method, which does not include wavelength dependent seeing but which does apply a correction to individual 15 minute exposures rather than to a summed one hour spectrum, works best. This overall pattern strongly implies (a) that it is indeed important to include wavelength dependant seeing in the pipeline SDSS procedures (not done at the time of writing this paper) and (b) that the DR16 method is likely to improve significantly when this is done.

\begin{table*}
  \begin{center}
    \caption{Comparing agreement between individual exposures for the quasar J022836.08+000939.2 (Thing ID 97176486) for each correction. The $\sigma$ on the second line of the header refers to the seeing profile i.e. Equation \eqref{eq:seeing}. The first two columns give the wavelength ranges over which the comparison statistic, $\xi^2$ (Eq.\eqref{eq:xisquared}), are calculated. The subsequent columns give $\xi^2$ and its estimated uncertainty, $\sigma(\xi^2)$, calculated from the number of degrees of freedom, Eq.\eqref{eq:sigxisq}. The number of degrees of freedom for the DR16 correction is very slightly different to the other cases but the difference is so small as to be negligible.}
    \label{tab:consistency_97176486}
    \begin{filecontents*}{tables/csv/97176486_variances.csv}
   3500.00,   4538.00,   7619.96,     37.07,   7598.39,     37.07,   7062.19,     37.07,   6915.53,     37.07,   7765.66,     36.33
   4538.00,   6500.00,  39286.45,     43.22,  14996.24,     43.22,  14016.92,     43.22,  12896.52,     43.22,  10906.41,     43.86
   6500.00,   8000.00,  66772.34,     29.83,  22643.63,     29.83,  22700.96,     29.83,  18748.33,     29.83,  14505.59,     29.87
   8000.00,  10400.00,  85504.20,     31.43,  30719.66,     31.43,  32314.00,     31.43,  24981.35,     31.43,  20376.50,     31.43
   3500.00,  10400.00, 199182.95,     71.89,  75957.92,     71.89,  76094.08,     71.89,  63541.73,     71.89,  53554.16,     71.92
\end{filecontents*}
\pgfplotstabletypeset[
      multicolumn names, % allows to have multicolumn names
      col sep=comma, % the separator in our .csv file
      every head row/.style=
        {
        before row={\toprule  \toprule
                     \multicolumn{2}{c}{Wavelength range} & 
                     \multicolumn{2}{c}{Uncorrected} & \multicolumn{2}{c}{Margala} & 
                     \multicolumn{2}{c}{Our correction} & \multicolumn{2}{c}{Our correction} & 
                     \multicolumn{2}{c}{DR16} \\
                     & & \multicolumn{2}{c}{} 
                     & \multicolumn{2}{c}{($\sigma=const.$)} & \multicolumn{2}{c}{($\sigma=const.$)} & \multicolumn{2}{c}{($\sigma=f(\lambda)$)} &
                     \multicolumn{2}{c}{($\sigma=const.$)} \\
                    },
        after row = {\midrule}
        }, 
        every last row/.style={after row=\bottomrule\bottomrule},
        display columns/0/.style=
          {
          column name = $\lambda_{min}$,
          precision=0,
          },
        display columns/1/.style=
          {
          column name=$\lambda_{max}$,
          precision=0,
          },
        display columns/2/.style=
          {
          column name=$\xi^2$,
          },
        display columns/3/.style=
          {
          column name=$\sigma(\xi^2)$,
          },
        display columns/4/.style=
          {
          column name=$\xi^2$,
          },
        display columns/5/.style=
          {
          column name=$\sigma(\xi^2)$,
          },
        display columns/6/.style=
          {
          column name=$\xi^2$,
          },
        display columns/7/.style=
          {
          column name=$\sigma(\xi^2)$,
          },
        display columns/8/.style=
          {
          column name=$\xi^2$,
          },
        display columns/9/.style=
          {
          column name=$\sigma(\xi^2)$,
          },
        display columns/10/.style=
          {
          column name=$\xi^2$,
          },
        display columns/11/.style=
          {
          column name=$\sigma(\xi^2)$,
          },
        fixed,
        fixed zerofill,
        precision=2,
        string replace={0.00}{},
        string replace={0}{},
        column type=r,
        set thousands separator = {\,},
        every nth row={4}{before row=\midrule}
    ]{tables/csv/97176486_variances.csv} % filename/path to file
  \end{center}
\end{table*}

\subsection{Spectral index measurements using different corrections}\label{sec:si}

\begin{figure*}
    \centering
    \includegraphics[width=\textwidth]{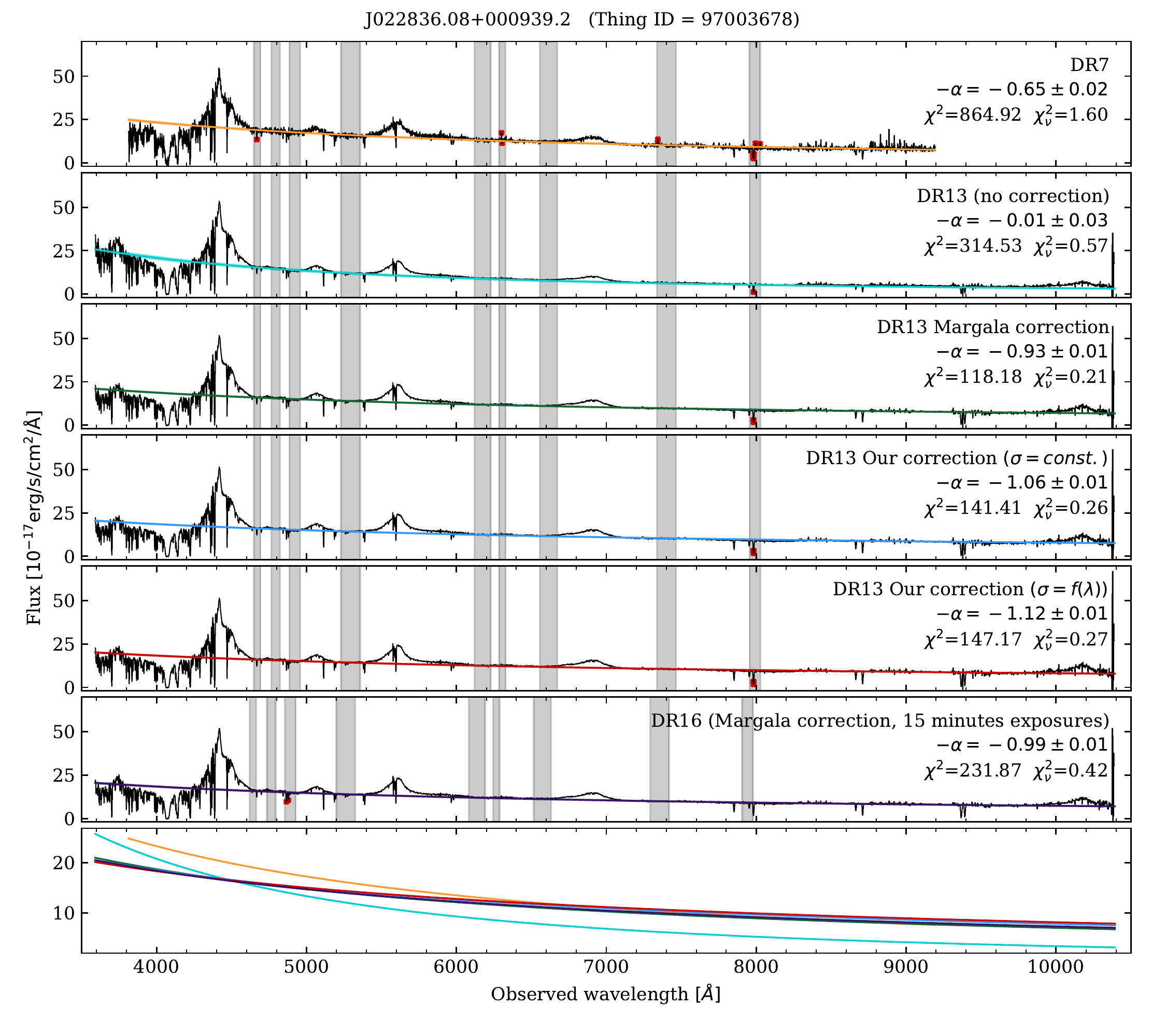}
    \caption{Summed spectrum of the quasar  J022836.08+000939.2 (black line). The top panel is the DR7 spectrum. The five spectra below the top panel are the same spectral data but corrected in different ways. The continuous coloured line in each panel shows the best-fit power law continuum to that spectrum, for different flux correction models. The text in each box shows the fitted spectral index and the best-fit $\chi^2$ (and its normalised value). The grey shaded areas illustrate the fitting regions. The red dots indicate pixels removed during the power law fitting procedure. See Section \ref{sec:si} for details. The six spectral panels all show that a single power law fits well. The lowest panel shows the six correction functions in slightly more detail.
    }
    \label{fig:97176486_mean_spectrum_comparison}
\end{figure*}

Another way in which we can explore consistency between different exposures of the same quasar is to compare spectral indices prior to and after applying a flux calibration correction. One would naturally expect greater consistency after the spectral shapes have been corrected. To compare spectral index measurements we first select nine small spectral regions that are minimally effected by quasar emission features. The rest-frame regions\footnote{\cite{Paris2017} also provide spectral indices for two of the four quasars we study here but use very different rest-frame fitting regions. Their regions are seen to contain emission line features in the composite quasar spectrum illustrated in figure\,4 of \cite{Zavarygin2019} so we avoid comparison with their results.} are: 1280-1292, 1312-1328, 1345-1365, 1440-1475, 1685-1715, 1730-1742, 1805-1837, 2020-2055, 2190-2210 \SI{}{\angstrom}, illustrated in \figref{fig:97176486_mean_spectrum_comparison}. We fit a linear function of the form
\be
\log_{10} f_\lambda = B + (\alpha-2) \log_{10} \lambda,
\label{eq:alpha}
\ee
where the specific flux $f_\lambda$ has the units erg s$^{-1}$ cm$^{-2}$ \AA$^{-1}$, $B$ is a constant, and (by convention) $-\alpha$ is the spectral index.  \figref{fig:97176486_mean_spectrum_comparison} shows that a single power law provides a good fit to the data at rest-frame wavelengths higher than approximately \SI{1280}{\angstrom}. Figure 4 of \cite{Zavarygin2019} illustrates this even more convincingly for a composite quasar spectrum. \figref{fig:97176486_mean_spectrum_comparison} illustrates the summed spectra for J022836.08+000939.2, for different flux calibration corrections. The best fit power law is shown in each case.
 
We use two approaches to minimise the impact from absorption features and other corrupted pixels falling within the fitting ranges on the spectral index determination. Firstly, we calculate the mean flux and the mean error in each fitting region and remove all pixels which are $\geq 3\times$ the mean error away from the mean flux in that region. We then fit the power-law of Equation~\eqref{eq:alpha} to the remaining data points, solving for $B$ and $\alpha$ by minimising the quantity
\be
\chi^2 = \sum_i \left( \frac{ f_i - f_i^c }{\sigma_i} \right)^2.
\label{eq:chisquared}
\ee
Here, the subscript $i$ corresponds to selected pixels within one spectrum, $f$ and $\sigma$ are the flux and its error (respectively), and $f^c$ is the estimated quasar continuum. We iteratively remove pixels for which the absolute values of the normalised residuals are $\geq 5$ until no more pixels are rejected. 

The mean spectral indices before and after fibre offset corrections are tabulated in Table \ref{tab:si_97176486}. Several interesting properties emerge. First, the standard deviation for all eight spectral index measurements is largest for the uncorrected measurements (0.42), see Table~\ref{tab:si_97176486}. All four corrections produce much greater consistency between the eight individually estimated spectral indices.

However, a scatter of 0.42 is much larger than the estimated uncertainty on the individual spectral index measurements. This tells us that either 022836.08+000939 undergoes intrinsic time variation or that the SDSS pipeline calibration leave significant residual uncertainties in the spectral shapes. We are unable to distinguish between these possibilities.

Tables \ref{tab:si_75018778} and \ref{tab:si_82763272} give the results for the second and third quasars and the same general result is seen; applying any of the corrections renders individual spectra more consistent but the scatter means we cannot favour one correction over another. We avoid the comparison for Table \ref{tab:si_110814178} for the reasons explained above. We learn little more by comparing the results with DR7. Of the three useful quasars, two have previous DR7 observations (unaffected by the fibre offset problem) which we use to measure spectral indices. For J022836.08+000939, $-\alpha_{\mathrm{DR7}} = -0.65 \pm 0.02$, which is inconsistent with any of the four methods shown in Table \ref{tab:si_97176486}. For J022954.42-005622, $-\alpha_{\mathrm{DR7}} = -0.96 \pm 0.03$. The Margala and DR16 values in the table are consistent but our two correction methods produce values that are too large. In summary, it turns out that this test is less informative than that described in Section \ref{sec:consistency} although it does indicate, in general terms, that the geometric correction methods work.

\begin{table*}
  \begin{center}
    \caption{Fitting power law continuum to individual one hour spectra for the quasar J022836.08+000939. The $\sigma$ on the second line of the header refers to the seeing profile i.e. Equation \eqref{eq:seeing}. The first column corresponds to the ID in Table \ref{tab:qso_exposures}. Columns 2 to 4 relate to uncorrected spectra. The second and the third columns give the best-fit spectral index, Equation \eqref{eq:alpha}, and its error ($\sigma_{\alpha}$). The fourth column is the best-fit normalised $\chi^2_{\nu}$, i.e. Equation \eqref{eq:chisquared} normalised by the number of degrees of freedom in the fit. The same pattern of three columns is repeated for each of the four correction methods. The last three lines in this table labelled $\mu$, $\sigma_{\mu}$, and $\sigma_{\mu}/\sqrt{8}$, give the unweighted mean spectral index, its standard deviation, and in the last line, the error on the mean. For comparison, the best-fit spectral index obtained (in the same way) using the SDSS DR7 spectrum is $-0.65\pm 0.02$. The mean value of $\chi^2_{\nu}$ gives an indication of the overall quality of fit.
    }
    \label{tab:si_97176486}
    \begin{filecontents*}{tables/csv/97176486.csv}
1    ,-0.59,0.01,1.32,-1.11,0.01,1.39,-1.27,0.01,1.44,-1.28,0.01,1.41,-1.11,0.01,1.51
2    ,-0.61,0.01,1.26,-1.03,0.01,1.34,-1.17,0.01,1.39,-1.22,0.01,1.36,-1.05,0.01,1.43
3    ,-0.58,0.01,1.64,-1.04,0.01,1.81,-1.17,0.01,1.92,-1.21,0.01,1.86,-1.08,0.01,2.01
4    ,+0.41,0.01,1.34,-0.57,0.01,1.49,-0.64,0.01,1.57,-0.72,0.01,1.48,-0.68,0.02,1.63
5    ,+0.08,0.01,1.11,-0.76,0.02,1.22,-0.94,0.02,1.30,-1.07,0.02,1.22,-0.96,0.01,1.14
6    ,+0.08,0.01,1.19,-0.73,0.01,1.28,-0.77,0.01,1.32,-0.86,0.01,1.27,-0.85,0.01,1.33
7    ,-0.17,0.01,0.97,-1.23,0.01,1.11,-1.26,0.01,1.15,-1.32,0.01,1.08,-1.15,0.02,1.25
8    ,+0.46,0.02,1.25,-0.45,0.02,1.35,-0.64,0.02,1.43,-0.76,0.02,1.36,-0.57,0.02,1.29
$\mu$,-0.11,0.00,1.26,-0.86,0.00,1.37,-0.98,0.00,1.44,-1.05,0.00,1.38,-0.93,0.00,1.45
$\sigma_\mu$,+0.42,0.00,0.00,+0.26,0.00,0.00,+0.25,0.00,0.00,+0.23,0.00,0.00,+0.20,0.00,0.00
$\sigma_\mu/\sqrt{8}$,+0.15,0.00,0.00,+0.09,0.00,0.00,+0.09,0.00,0.00,+0.08,0.00,0.00,+0.07,0.00,0.00
\end{filecontents*}
\pgfplotstabletypeset[
      multicolumn names, % allows to have multicolumn names
      col sep=comma, % the seperator in our .csv file
      every head row/.style=
        {
        before row={\toprule  \toprule
                     \multicolumn{1}{c}{\multirow{2}{*}{ID}} &  %\multicolumn{1}{c}{\multirow{2}{*}{Observation}} & 
                     \multicolumn{3}{c}{Uncorrected} & \multicolumn{3}{c}{Margala} & 
                     \multicolumn{3}{c}{Our correction} & \multicolumn{3}{c}{Our correction} & 
                     \multicolumn{3}{c}{DR16} \\
                     %& & \multicolumn{3}{c}{} 
                     & & & & \multicolumn{3}{c}{($\sigma=const.$)} & \multicolumn{3}{c}{($\sigma=const.$)} & \multicolumn{3}{c}{($\sigma=f(\lambda)$)} &
                     \multicolumn{3}{c}{($\sigma=const.$)} \\
                    },
        after row = {\midrule}
        }, 
        every last row/.style={after row=\bottomrule\bottomrule},
        display columns/0/.style=
          {
          column name = {},
          string type
          },
     %   display columns/1/.style=
      %    {
      %    column name=(Plate-MJD-Fibre),
      %    string type
      %    },
        display columns/1/.style=
          {
          column name=$-\alpha$,
          },
        display columns/2/.style=
          {
          column name=$\sigma_\alpha$,
          },
        display columns/3/.style=
          {
          column name=$\chi^2_\nu$,
          },
        display columns/4/.style=
          {
          column name=$-\alpha$,
          },
        display columns/5/.style=
          {
          column name=$\sigma_\alpha$,
          },
        display columns/6/.style=
          {
          column name=$\chi^2_\nu$,
          },
        display columns/7/.style=
          {
          column name=$-\alpha$,
          },
        display columns/8/.style=
          {
          column name=$\sigma_\alpha$,
          },
        display columns/9/.style=
          {
          column name=$\chi^2_\nu$,
          },
        display columns/10/.style=
          {
          column name=$-\alpha$,
          },
        display columns/11/.style=
          {
          column name=$\sigma_\alpha$,
          },
        display columns/12/.style=
          {
          column name=$\chi^2_\nu$,
          },
        display columns/13/.style=
          {
          column name=$-\alpha$,
          },
        display columns/14/.style=
          {
          column name=$\sigma_\alpha$,
          },
        display columns/15/.style=
          {
          column name=$\chi^2_\nu$,
          },
        fixed,
        fixed zerofill,
        precision=2,
        string replace={0.00}{},
        string replace={0}{},
        column type=r,
        every nth row={8}{before row=\midrule}
    ]{tables/csv/97176486.csv} % filename/path to file
  \end{center}
\end{table*}

\section{Discussion} \label{sec:discussion}

We have described a detailed geometric method for applying flux calibration corrections to SDSS quasar spectra, a problem first reported in \cite{Lee2013}. An earlier approach by \cite{Margala2016} is also geometric but differs from our in that we have used a correction in which the seeing is wavelength dependent. The results we obtain show that the earlier Margala correction works and also show that further improvement could be obtained for the default SDSS pipeline if wavelength dependent seeing is included. We note the two other independent flux correction methods of \cite{Harris2016, Guo2016} but since these are based on averaged stellar spectra and not a first-principles geometric approach, we have focused our comparisons purely between Margala and the new methods introduced in this paper. 

The calculations described here enable an assessment of the relative importance of individual corrections applied to solve the fibre offset problem. The left and right hand columns of \figref{fig:4params} provide a comparison between the four contributions to the overall flux corrections, without and with wavelength dependent seeing. It is clear that the functions do indeed change significantly, showing the importance of including wavelength dependent seeing.

We have compared four flux corrections, that of \cite{Margala2016}, our own two (one with and one without wavelength dependent seeing), and the DR16 correction (which is the Margala correction but applied individually to original 15 minute exposures rather than to grouped, one hour spectra). By examining the impact of these corrections on quasars with multiple exposures, we were able to see which correction method generated greater consistency between independent exposures. Comparing our own two methods, with and without wavelength dependent seeing, clearly illustrates the importance of allowing for the latter. 

Since our method was applied to grouped, one hour spectra, and the DR16 method applied to individual 15 minute exposures, the results detailed in Sections \ref{sec:consistency} and \ref{sec:si} strongly imply that the current DR16 correction method would improve substantially if wavelength dependent seeing were to be included.

Another point deserves mention although we have not investigated it in this paper. \cite{Zavarygin2019} show that there are spatially correlated systematic errors in the flux correction function over large scales across the sky. These systematics can emulate cosmological inhomogeneity. The origin of these effects in SDSS data is unknown but should be explored for certain science goals. Such effects should be avoided in any future similar surveys.

\section*{Acknowledgements}
JKW thanks the Department of Applied Mathematics and Theoretical Physics and the Institute of Astronomy at Cambridge University for hospitality and support, and Clare Hall for a Visiting Fellowship during part of this work. CCL thanks the Royal Society for a Newton International Fellowship during the early stages of this work.

\bibliographystyle{aa} % style aa.bst
\bibliography{offset} 

\begin{thebibliography}{30}
\expandafter\ifx\csname natexlab\endcsname\relax\def\natexlab#1{#1}\fi

\bibitem[{Ahn {et~al.}(2012)Ahn, Alexandroff, \& Prieto}]{Ahn2012}
Ahn, C.~P., Alexandroff, R., \& Prieto, C.~A. 2012, AJ, 203, 21

\bibitem[{{Ahumada} {et~al.}(2020){Ahumada}, {Prieto}, {Almeida}, {Anders},
  {Anderson}, {Andrews}, {Anguiano}, {Arcodia}, {Armengaud}, {Aubert}, {Avila},
  {Avila-Reese}, {Badenes}, {Balland}, {Barger}, {Barrera-Ballesteros}, {Basu},
  {Bautista}, {Beaton}, {Beers}, {Benavides}, {Bender}, {Bernardi}, {Bershady},
  {Beutler}, {Bidin}, {Bird}, {Bizyaev}, {Blanc}, {Blanton}, {Boquien},
  {Borissova}, {Bovy}, {Brandt}, {Brinkmann}, {Brownstein}, {Bundy}, {Bureau},
  {Burgasser}, {Burtin}, {Cano-D{\'\i}az}, {Capasso}, {Cappellari}, {Carrera},
  {Chabanier}, {Chaplin}, {Chapman}, {Cherinka}, {Chiappini}, {Doohyun Choi},
  {Chojnowski}, {Chung}, {Clerc}, {Coffey}, {Comerford}, {Comparat}, {da
  Costa}, {Cousinou}, {Covey}, {Crane}, {Cunha}, {Ilha}, {Dai}, {Damsted},
  {Darling}, {Davidson}, {Davies}, {Dawson}, {De}, {de la Macorra}, {De Lee},
  {Queiroz}, {Deconto Machado}, {de la Torre}, {Dell'Agli}, {du Mas des
  Bourboux}, {Diamond-Stanic}, {Dillon}, {Donor}, {Drory}, {Duckworth},
  {Dwelly}, {Ebelke}, {Eftekharzadeh}, {Davis Eigenbrot}, {Elsworth},
  {Eracleous}, {Erfanianfar}, {Escoffier}, {Fan}, {Farr},
  {Fern{\'a}ndez-Trincado}, {Feuillet}, {Finoguenov}, {Fofie},
  {Fraser-McKelvie}, {Frinchaboy}, {Fromenteau}, {Fu}, {Galbany}, {Garcia},
  {Garc{\'\i}a-Hern{\'a}ndez}, {Oehmichen}, {Ge}, {Maia}, {Geisler}, {Gelfand},
  {Goddy}, {Gonzalez-Perez}, {Grabowski}, {Green}, {Grier}, {Guo}, {Guy},
  {Harding}, {Hasselquist}, {Hawken}, {Hayes}, {Hearty}, {Hekker}, {Hogg},
  {Holtzman}, {Horta}, {Hou}, {Hsieh}, {Huber}, {Hunt}, {Chitham}, {Imig},
  {Jaber}, {Angel}, {Johnson}, {Jones}, {J{\"o}nsson}, {Jullo}, {Kim},
  {Kinemuchi}, {Kirkpatrick}, {Kite}, {Klaene}, {Kneib}, {Kollmeier}, {Kong},
  {Kounkel}, {Krishnarao}, {Lacerna}, {Lan}, {Lane}, {Law}, {Le Goff}, {Leung},
  {Lewis}, {Li}, {Lian}, {Lin}, {Long}, {Longa-Pe{\~n}a}, {Lundgren}, {Lyke},
  {Ted Mackereth}, {MacLeod}, {Majewski}, {Manchado}, {Maraston}, {Martini},
  {Masseron}, {Masters}, {Mathur}, {McDermid}, {Merloni}, {Merrifield},
  {M{\'e}sz{\'a}ros}, {Miglio}, {Minniti}, {Minsley}, {Miyaji}, {Mohammad},
  {Mosser}, {Mueller}, {Muna}, {Mu{\~n}oz-Guti{\'e}rrez}, {Myers}, {Nadathur},
  {Nair}, {Nandra}, {do Nascimento}, {Nevin}, {Newman}, {Nidever}, {Nitschelm},
  {Noterdaeme}, {O'Connell}, {Olmstead}, {Oravetz}, {Oravetz}, {Osorio},
  {Pace}, {Padilla}, {Palanque-Delabrouille}, {Palicio}, {Pan}, {Pan},
  {Parker}, {Paviot}, {Peirani}, {Ram{\'r}ez}, {Penny}, {Percival},
  {Perez-Fournon}, {P{\'e}rez-R{\`a}fols}, {Petitjean}, {Pieri},
  {Pinsonneault}, {Poovelil}, {Povick}, {Prakash}, {Price-Whelan}, {Raddick},
  {Raichoor}, {Ray}, {Rembold}, {Rezaie}, {Riffel}, {Riffel}, {Rix}, {Robin},
  {Roman-Lopes}, {Rom{\'a}n-Z{\'u}{\~n}iga}, {Rose}, {Ross}, {Rossi},
  {Rowlands}, {Rubin}, {Salvato}, {S{\'a}nchez}, {S{\'a}nchez-Menguiano},
  {S{\'a}nchez-Gallego}, {Sayres}, {Schaefer}, {Schiavon}, {Schimoia},
  {Schlafly}, {Schlegel}, {Schneider}, {Schultheis}, {Schwope}, {Seo},
  {Serenelli}, {Shafieloo}, {Shamsi}, {Shao}, {Shen}, {Shetrone}, {Shirley},
  {Aguirre}, {Simon}, {Skrutskie}, {Slosar}, {Smethurst}, {Sobeck}, {Sodi},
  {Souto}, {Stark}, {Stassun}, {Steinmetz}, {Stello}, {Stermer},
  {Storchi-Bergmann}, {Streblyanska}, {Stringfellow}, {Stutz}, {Su{\'a}rez},
  {Sun}, {Taghizadeh-Popp}, {Talbot}, {Tayar}, {Thakar}, {Theriault}, {Thomas},
  {Thomas}, {Tinker}, {Tojeiro}, {Toledo}, {Tremonti}, {Troup}, {Tuttle},
  {Unda-Sanzana}, {Valentini}, {Vargas-Gonz{\'a}lez}, {Vargas-Maga{\~n}a},
  {V{\'a}zquez-Mata}, {Vivek}, {Wake}, {Wang}, {Weaver}, {Weijmans}, {Wild},
  {Wilson}, {Wilson}, {Wolthuis}, {Wood-Vasey}, {Yan}, {Yang}, {Y{\`e}che},
  {Zamora}, {Zarrouk}, {Zasowski}, {Zhang}, {Zhao}, {Zhao}, {Zheng}, {Zheng},
  {Zhu}, \& {Zou}}]{Ahumada2020}
{Ahumada}, R., {Prieto}, C.~A., {Almeida}, A., {et~al.} 2020, \apjs, 249, 3

\bibitem[{{Alam} {et~al.}(2015){Alam}, {Albareti}, {Allende Prieto}, {Anders},
  {Anderson}, {Anderton}, {Andrews}, {Armengaud}, {Aubourg}, {Bailey}, \&
  et~al.}]{Alam2015}
{Alam}, S., {Albareti}, F.~D., {Allende Prieto}, C., {et~al.} 2015, \apjs, 219,
  12

\bibitem[{{Barnes} \& {Walsh}(1988)}]{Barnes1988}
{Barnes}, N.~P. \& {Walsh}, P.~J. 1988, AO, 27, 1230

\bibitem[{Buck(1981)}]{Buck1981}
Buck, A.~L. 1981, Journal of Applied Meteorology (1962-1982), 20, 1527

\bibitem[{{Dawson} {et~al.}(2013){Dawson}, {Schlegel}, {Ahn}, {Anderson},
  {Aubourg}, {Bailey}, {Barkhouser}, {Bautista}, {Beifiori}, {Berlind},
  {Bhardwaj}, {Bizyaev}, {Blake}, {Blanton}, {Blomqvist}, {Bolton}, {Borde},
  {Bovy}, {Brandt}, {Brewington}, {Brinkmann}, {Brown}, {Brownstein}, {Bundy},
  {Busca}, {Carithers}, {Carnero}, {Carr}, {Chen}, {Comparat}, {Connolly},
  {Cope}, {Croft}, {Cuesta}, {da Costa}, {Davenport}, {Delubac}, {de Putter},
  {Dhital}, {Ealet}, {Ebelke}, {Eisenstein}, {Escoffier}, {Fan}, {Filiz Ak},
  {Finley}, {Font-Ribera}, {G{\'e}nova-Santos}, {Gunn}, {Guo}, {Haggard},
  {Hall}, {Hamilton}, {Harris}, {Harris}, {Ho}, {Hogg}, {Holder}, {Honscheid},
  {Huehnerhoff}, {Jordan}, {Jordan}, {Kauffmann}, {Kazin}, {Kirkby}, {Klaene},
  {Kneib}, {Le Goff}, {Lee}, {Long}, {Loomis}, {Lundgren}, {Lupton}, {Maia},
  {Makler}, {Malanushenko}, {Malanushenko}, {Mandelbaum}, {Manera}, {Maraston},
  {Margala}, {Masters}, {McBride}, {McDonald}, {McGreer}, {McMahon}, {Mena},
  {Miralda-Escud{\'e}}, {Montero-Dorta}, {Montesano}, {Muna}, {Myers},
  {Naugle}, {Nichol}, {Noterdaeme}, {Nuza}, {Olmstead}, {Oravetz}, {Oravetz},
  {Owen}, {Padmanabhan}, {Palanque-Delabrouille}, {Pan}, {Parejko},
  {P{\^a}ris}, {Percival}, {P{\'e}rez-Fournon}, {P{\'e}rez-R{\`a}fols},
  {Petitjean}, {Pfaffenberger}, {Pforr}, {Pieri}, {Prada}, {Price-Whelan},
  {Raddick}, {Rebolo}, {Rich}, {Richards}, {Rockosi}, {Roe}, {Ross}, {Ross},
  {Rossi}, {Rubi{\~n}o-Martin}, {Samushia}, {S{\'a}nchez}, {Sayres}, {Schmidt},
  {Schneider}, {Sc{\'o}ccola}, {Seo}, {Shelden}, {Sheldon}, {Shen}, {Shu},
  {Slosar}, {Smee}, {Snedden}, {Stauffer}, {Steele}, {Strauss}, {Streblyanska},
  {Suzuki}, {Swanson}, {Tal}, {Tanaka}, {Thomas}, {Tinker}, {Tojeiro},
  {Tremonti}, {Vargas Maga{\~n}a}, {Verde}, {Viel}, {Wake}, {Watson}, {Weaver},
  {Weinberg}, {Weiner}, {West}, {White}, {Wood-Vasey}, {Yeche}, {Zehavi},
  {Zhao}, \& {Zheng}}]{Dawson2013}
{Dawson}, K.~S., {Schlegel}, D.~J., {Ahn}, C.~P., {et~al.} 2013, AJ, 145, 10

\bibitem[{Eisenstein {et~al.}(2011)Eisenstein, Weinberg, Agol, Aihara,
  Allende~Prieto, Anderson, \& Arns}]{Eisenstein2011}
Eisenstein, D.~J., Weinberg, D.~H., Agol, E., {et~al.} 2011, AJ, 142, 72

\bibitem[{{Filippenko}(1982)}]{Filippenko1982}
{Filippenko}, A.~V. 1982, PASP, 94, 715

\bibitem[{Gunn {et~al.}(2006)Gunn, Siegmund, Mannery, Owen, Hull, Leger, Carey,
  Knapp, York, Boroski, Kent, Lupton, Rockosi, Evans, Waddell, Anderson, Annis,
  Barentine, Bartoszek, Bastian, Bracker, Brewington, Briegel, Brinkmann,
  Brown, Carr, Czarapata, Drennan, Dombeck, Federwitz, Gillespie, Gonzales,
  Hansen, Harvanek, Hayes, Jordan, Kinney, Klaene, Kleinman, Kron, Kresinski,
  Lee, Limmongkol, Lindenmeyer, Long, Loomis, McGehee, Mantsch, Eric H~Neilsen,
  Neswold, Newman, Nitta, John~Peoples, Pier, Prieto, Prosapio, Rivetta,
  Schneider, Snedden, \& Wang}]{Gunn2006}
Gunn, J.~E., Siegmund, W.~A., Mannery, E.~J., {et~al.} 2006, AJ, 131, 2332

\bibitem[{{Guo} \& {Gu}(2016)}]{Guo2016}
{Guo}, H. \& {Gu}, M. 2016, \apj, 822, 26

\bibitem[{{Hanuschik}(2003)}]{Hanuschik2003}
{Hanuschik}, R.~W. 2003, \aap, 407, 1157

\bibitem[{{Harris} {et~al.}(2016){Harris}, {Jensen}, {Suzuki}, {Bautista},
  {Dawson}, {Vivek}, {Brownstein}, {Ge}, {Hamann}, {Herbst}, {Jiang}, {Moran},
  {Myers}, {Olmstead}, \& {Schneider}}]{Harris2016}
{Harris}, D.~W., {Jensen}, T.~W., {Suzuki}, N., {et~al.} 2016, \aj, 151, 155

\bibitem[{Kim {et~al.}(2002)Kim, Carswell, Cristiani, D'Odorico, \&
  Giallongo}]{Kim2002}
Kim, T.~S., Carswell, R.~F., Cristiani, S., D'Odorico, S., \& Giallongo, E.
  2002, \mnras, 335, 555

\bibitem[{{Lee} {et~al.}(2013){Lee}, {Bailey}, {Bartsch}, {Carithers},
  {Dawson}, {Kirkby}, {Lundgren}, {Margala}, {Palanque-Delabrouille}, {Pieri},
  {Schlegel}, {Weinberg}, {Y{\`e}che}, {Aubourg}, {Bautista}, {Bizyaev},
  {Blomqvist}, {Bolton}, {Borde}, {Brewington}, {Busca}, {Croft}, {Delubac},
  {Ebelke}, {Eisenstein}, {Font-Ribera}, {Ge}, {Hamilton}, {Hennawi}, {Ho},
  {Honscheid}, {Le Goff}, {Malanushenko}, {Malanushenko}, {Miralda-Escud{\'e}},
  {Myers}, {Noterdaeme}, {Oravetz}, {Pan}, {P{\^a}ris}, {Petitjean}, {Rich},
  {Rollinde}, {Ross}, {Rossi}, {Schneider}, {Simmons}, {Snedden}, {Slosar},
  {Spergel}, {Suzuki}, {Viel}, \& {Weaver}}]{Lee2013}
{Lee}, K.-G., {Bailey}, S., {Bartsch}, L.~E., {et~al.} 2013, \aj, 145, 69

\bibitem[{{Lehner} {et~al.}(2007){Lehner}, {Savage}, {Richter}, {Sembach},
  {Tripp}, \& {Wakker}}]{Lehner2007}
{Lehner}, N., {Savage}, B.~D., {Richter}, P., {et~al.} 2007, \apj, 658, 680

\bibitem[{{Lynds}(1971)}]{Lynds1971}
{Lynds}, R. 1971, \apjl, 164, L73

\bibitem[{{Margala} {et~al.}(2016){Margala}, {Kirkby}, {Dawson}, {Bailey},
  {Blanton}, \& {Schneider}}]{Margala2016}
{Margala}, D., {Kirkby}, D., {Dawson}, K., {et~al.} 2016, \apj, 831, 157

\bibitem[{{Marini} \& {Murray}(1973)}]{Marini1973}
{Marini}, J.~W. \& {Murray}, C.~W. 1973, NASA Technical Report X-591-73-351

\bibitem[{{Meyers} \& {Burchat}(2015)}]{Meyers2015}
{Meyers}, J.~E. \& {Burchat}, P.~R. 2015, \apj, 807, 182

\bibitem[{{P{\^a}ris} {et~al.}(2017){P{\^a}ris}, {Petitjean}, {Ross}, {Myers},
  {Aubourg}, {Streblyanska}, {Bailey}, {Armengaud}, {Palanque-Delabrouille},
  {Y{\`e}che}, {Hamann}, {Strauss}, {Albareti}, {Bovy}, {Bizyaev}, {Niel
  Brandt}, {Brusa}, {Buchner}, {Comparat}, {Croft}, {Dwelly}, {Fan},
  {Font-Ribera}, {Ge}, {Georgakakis}, {Hall}, {Jiang}, {Kinemuchi},
  {Malanushenko}, {Malanushenko}, {McMahon}, {Menzel}, {Merloni}, {Nandra},
  {Noterdaeme}, {Oravetz}, {Pan}, {Pieri}, {Prada}, {Salvato}, {Schlegel},
  {Schneider}, {Simmons}, {Viel}, {Weinberg}, \& {Zhu}}]{Paris2017}
{P{\^a}ris}, I., {Petitjean}, P., {Ross}, N.~P., {et~al.} 2017, \aap, 597, A79

\bibitem[{Penton {et~al.}(2004)Penton, Stocke, \& Shull}]{Penton2004}
Penton, S.~V., Stocke, J.~T., \& Shull, J.~M. 2004, \apjs, 152, 29

\bibitem[{{Rauch}(1998)}]{Rauch1998}
{Rauch}, M. 1998, \araa, 36, 267

\bibitem[{{Roddier}(1981)}]{Roddier1981}
{Roddier}, F. 1981, Progess in Optics, 19, 281

\bibitem[{Ross {et~al.}(2012)Ross, Myers, Sheldon, Y{\`e}che, Strauss, Bovy,
  Kirkpatrick, Richards, Aubourg, Blanton, Brandt, Carithers, Croft, da~Silva,
  Dawson, Eisenstein, Hennawi, Ho, Hogg, Lee, Lundgren, McMahon,
  Miralda-Escud{\'e}, Palanque-Delabrouille, Paris, Petitjean, Pieri, Rich,
  Roe, Schiminovich, Schlegel, Schneider, Slosar, Suzuki, Tinker, Weinberg,
  Weyant, White, \& Wood-Vasey}]{Ross2012}
Ross, N.~P., Myers, A.~D., Sheldon, E.~S., {et~al.} 2012, \apjs, 199, 3

\bibitem[{{Sargent} {et~al.}(1980){Sargent}, {Young}, {Boksenberg}, \&
  {Tytler}}]{Sargent1980}
{Sargent}, W.~L.~W., {Young}, P.~J., {Boksenberg}, A., \& {Tytler}, D. 1980,
  \apjs, 42, 41

\bibitem[{{Smart}(1931)}]{Smart1931}
{Smart}, W.~M. 1931, {Spherical astronomy} (Cambridge [Eng.] The University
  press, 1931.)

\bibitem[{Smee {et~al.}(2013)Smee, Gunn, Uomoto, Roe, Schlegel, \&
  Rockosi}]{Smee2013}
Smee, S.~A., Gunn, J.~E., Uomoto, A., {et~al.} 2013, AJ, 146, 32

\bibitem[{{Weymann} {et~al.}(1981){Weymann}, {Carswell}, \&
  {Smith}}]{Weymann1981}
{Weymann}, R.~J., {Carswell}, R.~F., \& {Smith}, M.~G. 1981, \araa, 19, 41

\bibitem[{{York} {et~al.}(2000){York}, {Adelman}, {Anderson}, {Anderson},
  {Annis}, {Bahcall}, {Bakken}, {Barkhouser}, {Bastian}, {Berman}, {Boroski},
  {Bracker}, {Briegel}, {Briggs}, {Brinkmann}, {Brunner}, {Burles}, {Carey},
  {Carr}, {Castander}, {Chen}, {Colestock}, {Connolly}, {Crocker}, {Csabai},
  {Czarapata}, {Davis}, {Doi}, {Dombeck}, {Eisenstein}, {Ellman}, {Elms},
  {Evans}, {Fan}, {Federwitz}, {Fiscelli}, {Friedman}, {Frieman}, {Fukugita},
  {Gillespie}, {Gunn}, {Gurbani}, {de Haas}, {Haldeman}, {Harris}, {Hayes},
  {Heckman}, {Hennessy}, {Hindsley}, {Holm}, {Holmgren}, {Huang}, {Hull},
  {Husby}, {Ichikawa}, {Ichikawa}, {Ivezi{\'c}}, {Kent}, {Kim}, {Kinney},
  {Klaene}, {Kleinman}, {Kleinman}, {Knapp}, {Korienek}, {Kron}, {Kunszt},
  {Lamb}, {Lee}, {Leger}, {Limmongkol}, {Lindenmeyer}, {Long}, {Loomis},
  {Loveday}, {Lucinio}, {Lupton}, {MacKinnon}, {Mannery}, {Mantsch}, {Margon},
  {McGehee}, {McKay}, {Meiksin}, {Merelli}, {Monet}, {Munn}, {Narayanan},
  {Nash}, {Neilsen}, {Neswold}, {Newberg}, {Nichol}, {Nicinski}, {Nonino},
  {Okada}, {Okamura}, {Ostriker}, {Owen}, {Pauls}, {Peoples}, {Peterson},
  {Petravick}, {Pier}, {Pope}, {Pordes}, {Prosapio}, {Rechenmacher}, {Quinn},
  {Richards}, {Richmond}, {Rivetta}, {Rockosi}, {Ruthmansdorfer}, {Sand ford},
  {Schlegel}, {Schneider}, {Sekiguchi}, {Sergey}, {Shimasaku}, {Siegmund},
  {Smee}, {Smith}, {Snedden}, {Stone}, {Stoughton}, {Strauss}, {Stubbs},
  {SubbaRao}, {Szalay}, {Szapudi}, {Szokoly}, {Thakar}, {Tremonti}, {Tucker},
  {Uomoto}, {Vanden Berk}, {Vogeley}, {Waddell}, {Wang}, {Watanabe},
  {Weinberg}, {Yanny}, {Yasuda}, \& {SDSS Collaboration}}]{York2000}
{York}, D.~G., {Adelman}, J., {Anderson}, John~E., J., {et~al.} 2000, \aj, 120,
  1579

\bibitem[{{Zavarygin} \& {Webb}(2019)}]{Zavarygin2019}
{Zavarygin}, E.~O. \& {Webb}, J.~K. 2019, \mnras, 489, 3966

\end{thebibliography}

\clearpage

%=====================================================

\appendix

\section{Second quasar details, J022954.42-005622.5   (Thing ID = 74944092)} \label{sec:appendixA}

\begin{figure*}
    \centering
    \includegraphics[width=\textwidth]{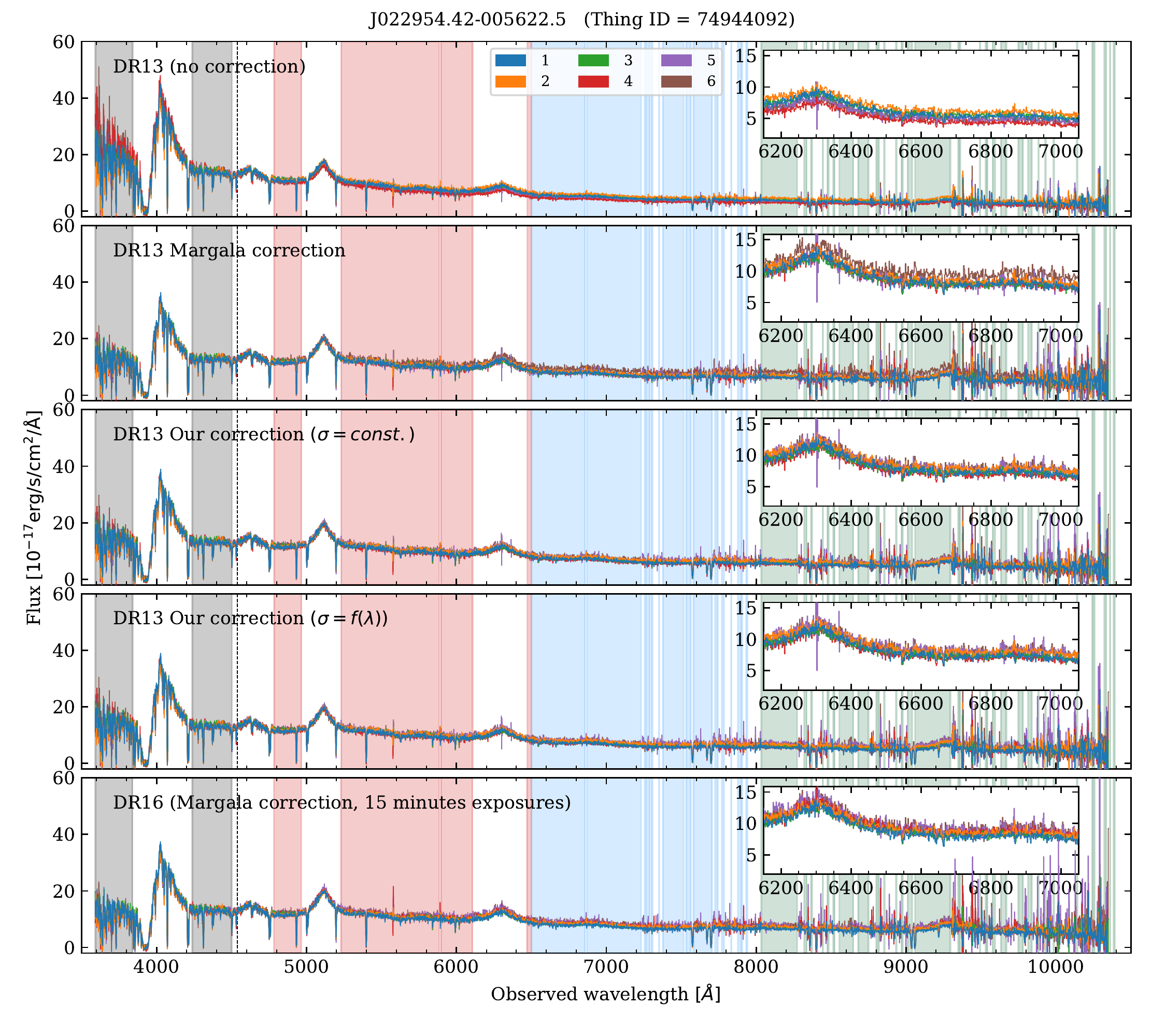}
    \caption{Same as \figref{fig:spec_obs} except for the quasar J022954.42-005622.5.}
    \label{fig:J023308corrections2}
\end{figure*}

\begin{figure*}
    \centering
    \includegraphics[width=\textwidth]{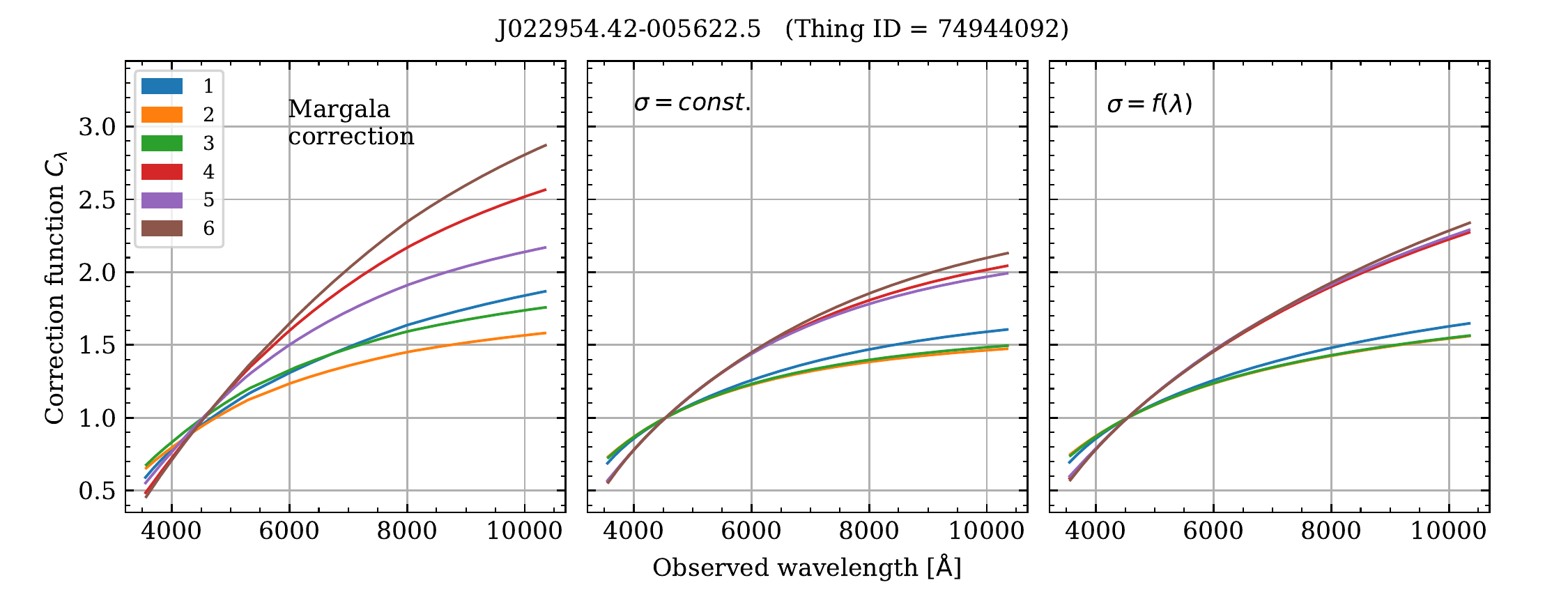}
    \caption{Same as \figref{fig:corr} except for the quasar J022954.42-005622.5.}
    \label{fig:75018778_calib_corr_comparison}
\end{figure*}

\begin{figure*}
    \centering
    \includegraphics[width=\textwidth]{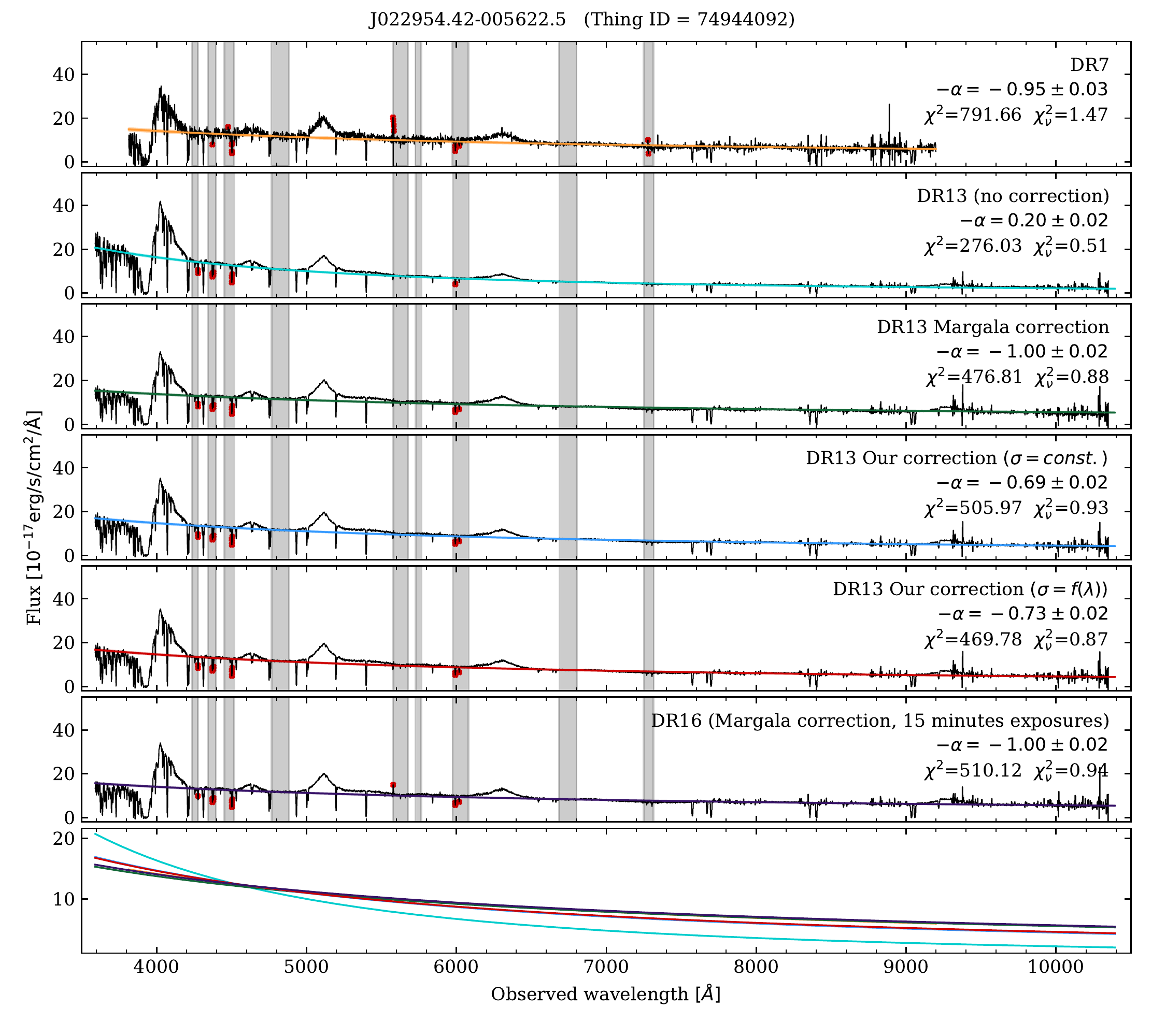}
    \caption{Same as \figref{fig:97176486_mean_spectrum_comparison} except for the quasar J022954.42-005622.5.}
    \label{fig:J023308corrections1}
\end{figure*}

\begin{table*}
  \begin{center}
    \caption{Same as Table \ref{tab:consistency_97176486} except for the quasar J022954.42-005622.5.
%Comparing agreement between individual exposures for the SDSS quasar J022954.42-005622.5 (Thing ID 75018778) for each correction. The first two columns give the wavelength ranges over which the comparison statistic, $\xi^2$ (Eq.\eqref{eq:xisquared}) are made. The subsequent columns give $\xi^2$ and its estimated uncertainty, $\sigma(\xi^2)$, calculated from the number of degrees of freedom, Eq.\eqref{eq:sigxisq}.
}
    \label{tab:consistency_75018778}
    \begin{filecontents*}{tables/csv/75018778_variances.csv}
   3500.00,   4538.00,   4374.39,     33.14,   3176.05,     33.14,   3259.47,     33.14,   3236.89,     33.14,   3253.65,     33.11
   4538.00,   6500.00,  10560.97,     41.04,   4324.04,     41.04,   4403.58,     41.04,   4456.87,     41.04,   4455.37,     41.06
   6500.00,   8000.00,  20151.37,     37.20,   6332.29,     37.20,   6233.58,     37.20,   6754.04,     37.20,   5497.51,     37.20
   8000.00,  10400.00,  14229.13,     31.50,   4558.46,     31.50,   4705.65,     31.50,   5611.44,     31.50,   3835.23,     31.50
   3500.00,  10400.00,  49315.85,     72.07,  18390.84,     72.07,  18602.28,     72.07,  20059.24,     72.07,  17041.76,     72.07
\end{filecontents*}
\pgfplotstabletypeset[
      multicolumn names, % allows to have multicolumn names
      col sep=comma, % the separator in our .csv file
      every head row/.style=
        {
        before row={\toprule  \toprule
                     \multicolumn{2}{c}{Wavelength range} & 
                     \multicolumn{2}{c}{Uncorrected} & \multicolumn{2}{c}{Margala} & 
                     \multicolumn{2}{c}{Our correction} & \multicolumn{2}{c}{Our correction} & 
                     \multicolumn{2}{c}{DR16} \\
                     & & \multicolumn{2}{c}{} 
                     & \multicolumn{2}{c}{($\sigma=const.$)} & \multicolumn{2}{c}{($\sigma=const.$)} & \multicolumn{2}{c}{($\sigma=f(\lambda)$)} &
                     \multicolumn{2}{c}{($\sigma=const.$)} \\
                    },
        after row = {\midrule}
        }, 
        every last row/.style={after row=\bottomrule\bottomrule},
        display columns/0/.style=
          {
          column name = $\lambda_{min}$,
          precision=0,
          },
        display columns/1/.style=
          {
          column name=$\lambda_{max}$,
          precision=0,
          },
        display columns/2/.style=
          {
          column name=$\xi^2$,
          },
        display columns/3/.style=
          {
          column name=$\sigma(\xi^2)$,
          },
        display columns/4/.style=
          {
          column name=$\xi^2$,
          },
        display columns/5/.style=
          {
          column name=$\sigma(\xi^2)$,
          },
        display columns/6/.style=
          {
          column name=$\xi^2$,
          },
        display columns/7/.style=
          {
          column name=$\sigma(\xi^2)$,
          },
        display columns/8/.style=
          {
          column name=$\xi^2$,
          },
        display columns/9/.style=
          {
          column name=$\sigma(\xi^2)$,
          },
        display columns/10/.style=
          {
          column name=$\xi^2$,
          },
        display columns/11/.style=
          {
          column name=$\sigma(\xi^2)$,
          },
        fixed,
        fixed zerofill,
        precision=2,
        string replace={0.00}{},
        string replace={0}{},
        column type=r,
        set thousands separator = {\,},
        every nth row={4}{before row=\midrule}
    ]{tables/csv/75018778_variances.csv} % filename/path to file
  \end{center}
\end{table*}

\begin{table*}
  \begin{center}
    \caption{Same as Table \ref{tab:si_97176486} except for the quasar J022954.42-005622.5. The SDSS DR7 spectral index is $-0.96\pm0.03$.}
    \label{tab:si_75018778}
    \begin{filecontents*}{tables/csv/75018778.csv}
1    ,+0.13,0.02,1.40,-0.93,0.02,1.76,-0.65,0.02,1.66,-0.65,0.02,1.65,-0.89,0.02,1.62
2    ,-0.18,0.02,1.42,-1.04,0.02,1.93,-0.85,0.02,1.80,-0.88,0.02,1.72,-1.04,0.02,1.62
3    ,+0.11,0.01,1.55,-0.83,0.02,2.02,-0.58,0.02,1.89,-0.60,0.02,1.83,-0.84,0.02,1.80
4    ,+0.70,0.02,1.09,-0.88,0.02,1.66,-0.51,0.02,1.50,-0.57,0.02,1.39,-1.02,0.02,1.55
5    ,+0.45,0.02,1.15,-0.90,0.02,1.54,-0.77,0.02,1.49,-0.85,0.02,1.38,-1.12,0.02,1.46
6    ,+0.41,0.02,1.15,-1.32,0.02,1.44,-0.85,0.02,1.35,-0.89,0.02,1.31,-1.16,0.02,1.37
$\mu$,+0.27,0.00,1.29,-0.98,0.00,1.72,-0.70,0.00,1.62,-0.74,0.00,1.55,-1.01,0.00,1.57
$\sigma_\mu$,+0.29,0.00,0.00,+0.17,0.00,0.00,+0.13,0.00,0.00,+0.14,0.00,0.00,+0.12,0.00,0.00
$\sigma_\mu/\sqrt{6}$,+0.12,0.00,0.00,+0.07,0.00,0.00,+0.05,0.00,0.00,+0.06,0.00,0.00,+0.05,0.00,0.00
\end{filecontents*}
\pgfplotstabletypeset[
      multicolumn names, % allows to have multicolumn names
      col sep=comma, % the seperator in our .csv file
      every head row/.style=
        {
        before row={\toprule  \toprule
                     \multicolumn{1}{c}{\multirow{2}{*}{ID}} &  %\multicolumn{1}{c}{\multirow{2}{*}{Observation}} & 
                     \multicolumn{3}{c}{Uncorrected} & \multicolumn{3}{c}{Margala} & 
                     \multicolumn{3}{c}{Our correction} & \multicolumn{3}{c}{Our correction} & 
                     \multicolumn{3}{c}{DR16} \\
                     %& & \multicolumn{3}{c}{} 
                     & & & & \multicolumn{3}{c}{($\sigma=const.$)} & \multicolumn{3}{c}{($\sigma=const.$)} & \multicolumn{3}{c}{($\sigma=f(\lambda)$)} &
                     \multicolumn{3}{c}{($\sigma=const.$)} \\
                    },
        after row = {\midrule}
        }, 
        every last row/.style={after row=\bottomrule\bottomrule},
        display columns/0/.style=
          {
          column name = {},
          string type
          },
     %   display columns/1/.style=
      %    {
      %    column name=(Plate-MJD-Fibre),
      %    string type
      %    },
        display columns/1/.style=
          {
          column name=$-\alpha$,
          },
        display columns/2/.style=
          {
          column name=$\sigma_\alpha$,
          },
        display columns/3/.style=
          {
          column name=$\chi^2_\nu$,
          },
        display columns/4/.style=
          {
          column name=$-\alpha$,
          },
        display columns/5/.style=
          {
          column name=$\sigma_\alpha$,
          },
        display columns/6/.style=
          {
          column name=$\chi^2_\nu$,
          },
        display columns/7/.style=
          {
          column name=$-\alpha$,
          },
        display columns/8/.style=
          {
          column name=$\sigma_\alpha$,
          },
        display columns/9/.style=
          {
          column name=$\chi^2_\nu$,
          },
        display columns/10/.style=
          {
          column name=$-\alpha$,
          },
        display columns/11/.style=
          {
          column name=$\sigma_\alpha$,
          },
        display columns/12/.style=
          {
          column name=$\chi^2_\nu$,
          },
        display columns/13/.style=
          {
          column name=$-\alpha$,
          },
        display columns/14/.style=
          {
          column name=$\sigma_\alpha$,
          },
        display columns/15/.style=
          {
          column name=$\chi^2_\nu$,
          },
        fixed,
        fixed zerofill,
        precision=2,
        string replace={0.00}{},
        string replace={0}{},
        column type=r,
        every nth row={6}{before row=\midrule}
    ]{tables/csv/75018778.csv} % filename/path to file
  \end{center}
\end{table*}

\clearpage

\section{Third quasar details, J023308.31-002605.0   (Thing ID = 82655414)} \label{sec:appendixB}

\begin{figure*}
    \centering
    \includegraphics[width=\textwidth]{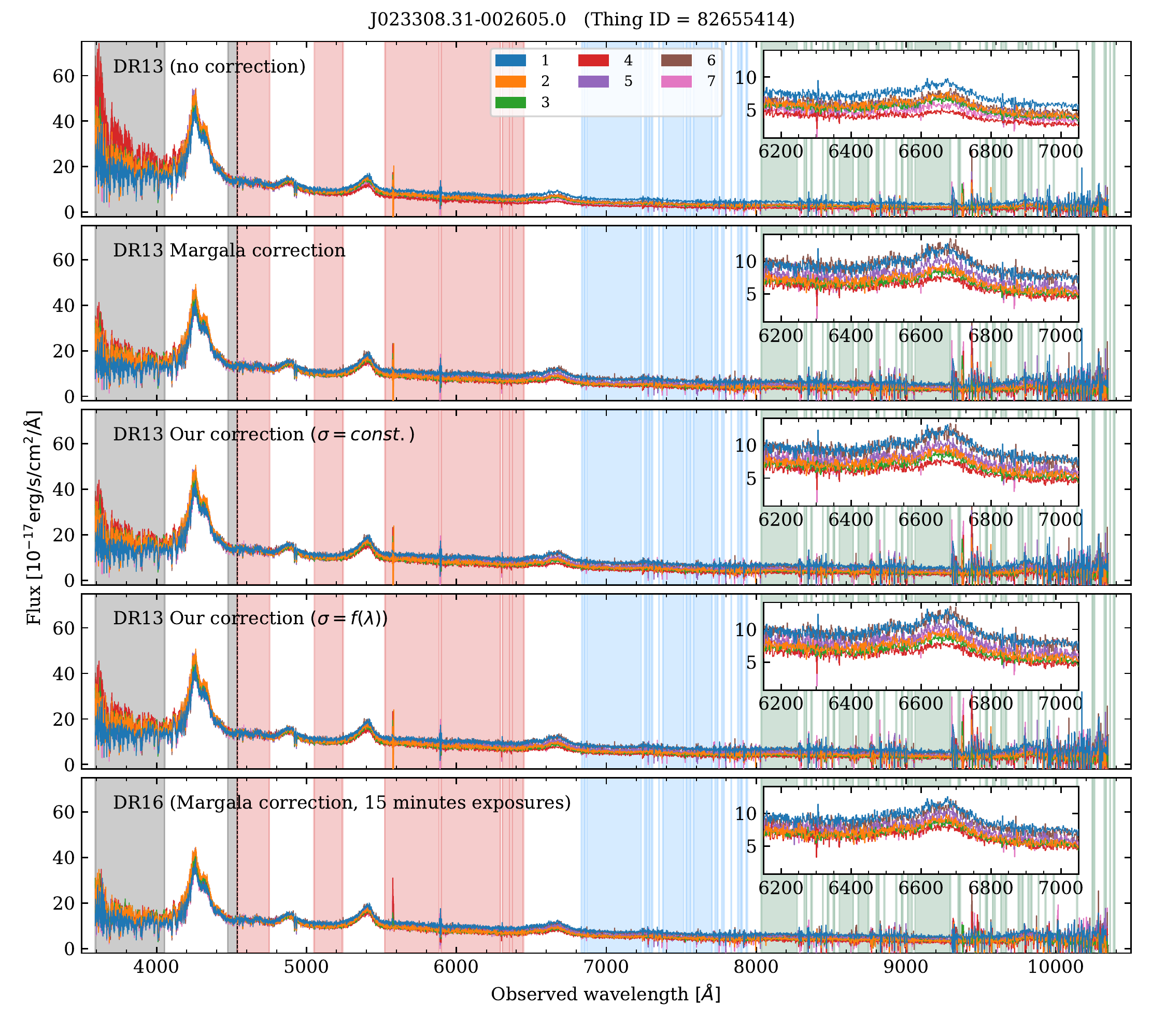}
    \caption{Same as \figref{fig:spec_obs} except for the quasar J023308.31-002605.0.}
    \label{fig:J023258correctionsB}
\end{figure*}

\begin{figure*}
    \centering
    \includegraphics[width=\textwidth]{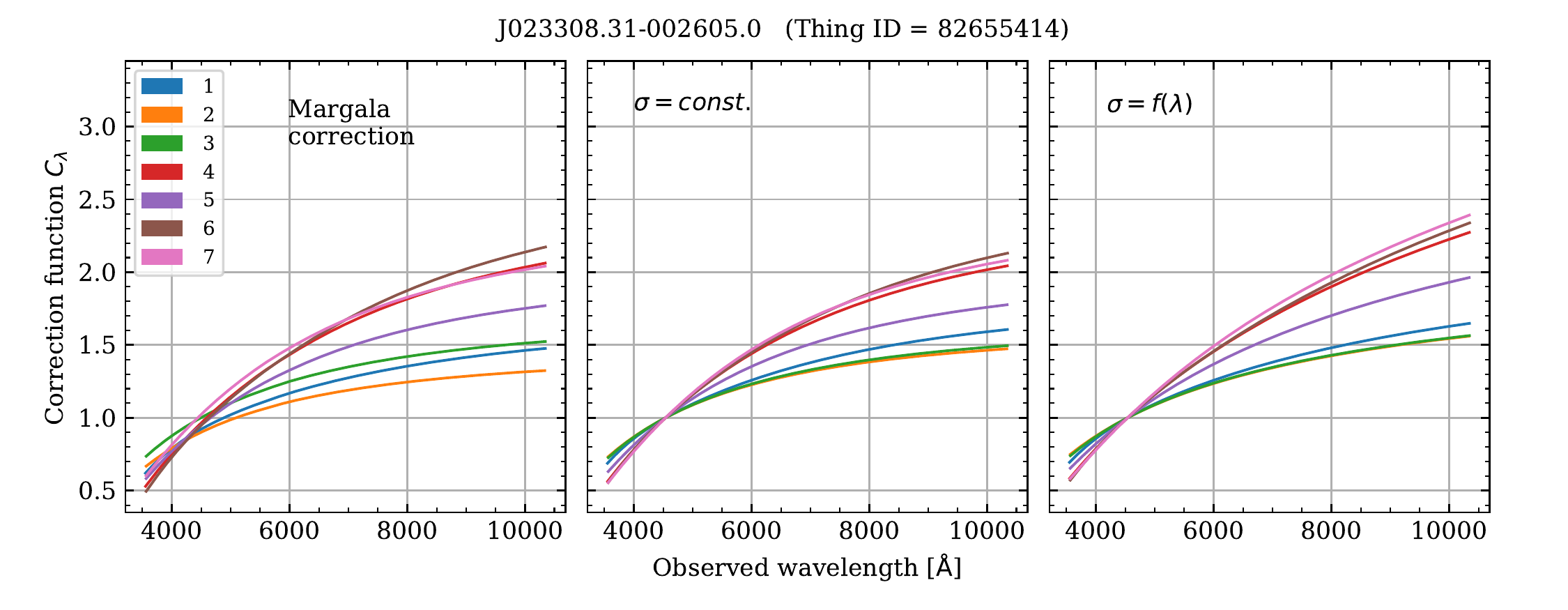}
    \caption{Same as \figref{fig:corr} except for the quasar J023308.31-002605.0.}
    \label{fig:82763272_calib_corr_comparison}
\end{figure*}

\begin{figure*}
    \centering
    \includegraphics[width=\textwidth]{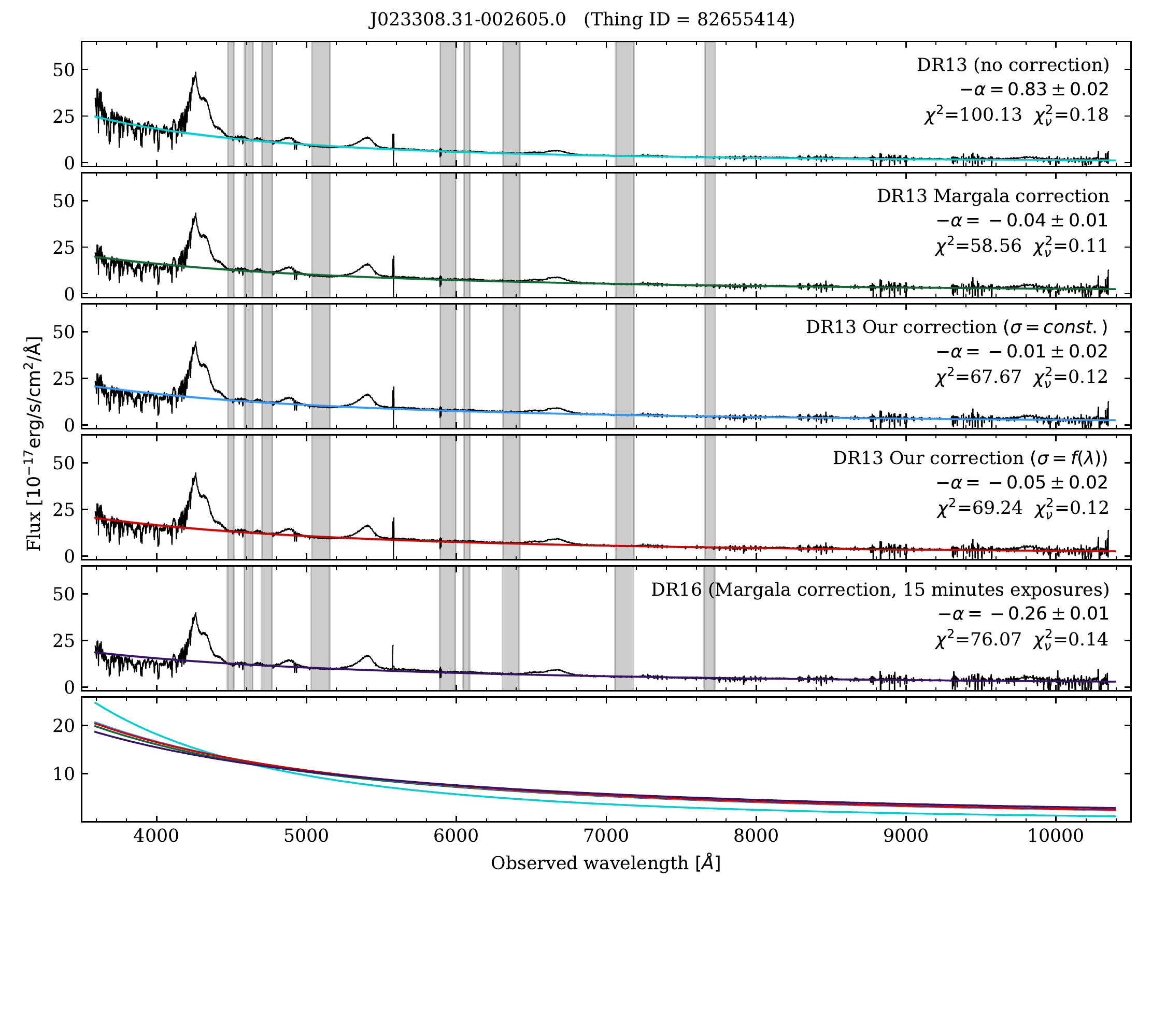}
    \caption{Same as \figref{fig:97176486_mean_spectrum_comparison} except for the quasar J023308.31-002605.0.}
    \label{fig:J022836corrections1}
\end{figure*}

\begin{table*}
  \begin{center}
    \caption{Same as Table \ref{tab:consistency_97176486} except for the quasar J023308.31-002605.0.
%Comparing agreement between individual exposures for the SDSS quasar J023308.31-002605.0 (Thing ID 82763272) for each correction. The first two columns give the wavelength ranges over which the comparison statistic, $\xi^2$ (Eq.\eqref{eq:xisquared}) are made. The subsequent columns give $\xi^2$ and its estimated uncertainty, $\sigma(\xi^2)$, calculated from the number of degrees of freedom, Eq.\eqref{eq:sigxisq}.
}
    \label{tab:consistency_827763272}
    \begin{filecontents*}{tables/csv/82763272_variances.csv}
   3500.00,   4538.00,  10875.62,     34.12,   7808.13,     34.12,   7885.58,     34.12,   8072.62,     34.12,   7719.01,     34.12
   4538.00,   6500.00,  27229.68,     44.92,  21459.65,     44.92,  21222.66,     44.92,  20710.85,     44.92,  16885.60,     44.83
   6500.00,   8000.00,  32276.55,     30.76,  25497.02,     30.76,  25754.19,     30.76,  24741.45,     30.76,  17308.30,     30.82
   8000.00,  10400.00,  31053.04,     31.46,  23403.65,     31.46,  23933.95,     31.46,  22658.79,     31.46,  16458.13,     31.46
   3500.00,  10400.00, 101434.88,     71.83,  78168.45,     71.83,  78796.38,     71.83,  76183.71,     71.83,  58371.04,     71.81
\end{filecontents*}
\pgfplotstabletypeset[
      multicolumn names, % allows to have multicolumn names
      col sep=comma, % the separator in our .csv file
      every head row/.style=
        {
        before row={\toprule  \toprule
                     \multicolumn{2}{c}{Wavelength range} & 
                     \multicolumn{2}{c}{Uncorrected} & \multicolumn{2}{c}{Margala} & 
                     \multicolumn{2}{c}{Our correction} & \multicolumn{2}{c}{Our correction} & 
                     \multicolumn{2}{c}{DR16} \\
                     & & \multicolumn{2}{c}{} 
                     & \multicolumn{2}{c}{($\sigma=const.$)} & \multicolumn{2}{c}{($\sigma=const.$)} & \multicolumn{2}{c}{($\sigma=f(\lambda)$)} &
                     \multicolumn{2}{c}{($\sigma=const.$)} \\
                    },
        after row = {\midrule}
        }, 
        every last row/.style={after row=\bottomrule\bottomrule},
        display columns/0/.style=
          {
          column name = $\lambda_{min}$,
          precision=0,
          },
        display columns/1/.style=
          {
          column name=$\lambda_{max}$,
          precision=0,
          },
        display columns/2/.style=
          {
          column name=$\xi^2$,
          },
        display columns/3/.style=
          {
          column name=$\sigma(\xi^2)$,
          },
        display columns/4/.style=
          {
          column name=$\xi^2$,
          },
        display columns/5/.style=
          {
          column name=$\sigma(\xi^2)$,
          },
        display columns/6/.style=
          {
          column name=$\xi^2$,
          },
        display columns/7/.style=
          {
          column name=$\sigma(\xi^2)$,
          },
        display columns/8/.style=
          {
          column name=$\xi^2$,
          },
        display columns/9/.style=
          {
          column name=$\sigma(\xi^2)$,
          },
        display columns/10/.style=
          {
          column name=$\xi^2$,
          },
        display columns/11/.style=
          {
          column name=$\sigma(\xi^2)$,
          },
        fixed,
        fixed zerofill,
        precision=2,
        string replace={0.00}{},
        string replace={0}{},
        column type=r,
        set thousands separator = {\,},
        every nth row={4}{before row=\midrule}
    ]{tables/csv/82763272_variances.csv} % filename/path to file
  \end{center}
\end{table*}

\begin{table*}
  \begin{center}
    \caption{Same as Table \ref{tab:si_97176486} except for the quasar J023308.31-002605.0. \mgt{(EZ: Give DR7 value)}}
    \label{tab:si_82763272}
    \begin{filecontents*}{tables/csv/82763272.csv}
1    ,-0.09,0.02,1.48,-0.78,0.02,1.55,-0.80,0.02,1.54,-0.81,0.02,1.52,-0.80,0.02,1.64
2    ,+0.63,0.02,1.17,+0.03,0.02,1.12,+0.01,0.02,1.12,-0.03,0.02,1.12,-0.09,0.02,1.07
3    ,+0.82,0.02,1.57,+0.17,0.02,1.40,+0.19,0.02,1.39,+0.15,0.02,1.40,-0.03,0.01,1.12
4    ,+1.58,0.02,1.40,+0.40,0.02,1.30,+0.45,0.02,1.31,+0.38,0.02,1.31,+0.13,0.02,1.23
5    ,+0.66,0.03,1.55,-0.31,0.02,1.35,-0.26,0.02,1.36,-0.33,0.02,1.39,-0.37,0.02,1.10
6    ,+0.50,0.02,1.32,-0.75,0.02,1.26,-0.66,0.02,1.26,-0.71,0.02,1.25,-0.61,0.02,1.15
7    ,+1.02,0.02,1.23,-0.07,0.02,1.12,-0.16,0.02,1.11,-0.26,0.02,1.12,-0.24,0.02,1.18
$\mu$,+0.73,0.00,1.39,-0.19,0.00,1.30,-0.18,0.00,1.30,-0.23,0.00,1.30,-0.29,0.00,1.21
$\sigma_\mu$,+0.47,0.00,0.00,+0.42,0.00,0.00,+0.41,0.00,0.00,+0.40,0.00,0.00,+0.31,0.00,0.00
$\sigma_\mu/\sqrt{7}$,+0.18,0.00,0.00,+0.16,0.00,0.00,+0.16,0.00,0.00,+0.15,0.00,0.00,+0.12,0.00,0.00
\end{filecontents*}
\pgfplotstabletypeset[
      multicolumn names, % allows to have multicolumn names
      col sep=comma, % the seperator in our .csv file
      every head row/.style=
        {
        before row={\toprule  \toprule
                     \multicolumn{1}{c}{\multirow{2}{*}{ID}} &  %\multicolumn{1}{c}{\multirow{2}{*}{Observation}} & 
                     \multicolumn{3}{c}{Uncorrected} & \multicolumn{3}{c}{Margala} & 
                     \multicolumn{3}{c}{Our correction} & \multicolumn{3}{c}{Our correction} & 
                     \multicolumn{3}{c}{DR16} \\
                     %& & \multicolumn{3}{c}{} 
                     & & & & \multicolumn{3}{c}{($\sigma=const.$)} & \multicolumn{3}{c}{($\sigma=const.$)} & \multicolumn{3}{c}{($\sigma=f(\lambda)$)} &
                     \multicolumn{3}{c}{($\sigma=const.$)} \\
                    },
        after row = {\midrule}
        },  
        every last row/.style={after row=\bottomrule\bottomrule},
        display columns/0/.style=
          {
          column name = {},
          string type
          },
     %   display columns/1/.style=
      %    {
      %    column name=(Plate-MJD-Fibre),
      %    string type
      %    },
        display columns/1/.style=
          {
          column name=$-\alpha$,
          },
        display columns/2/.style=
          {
          column name=$\sigma_\alpha$,
          },
        display columns/3/.style=
          {
          column name=$\chi^2_\nu$,
          },
        display columns/4/.style=
          {
          column name=$-\alpha$,
          },
        display columns/5/.style=
          {
          column name=$\sigma_\alpha$,
          },
        display columns/6/.style=
          {
          column name=$\chi^2_\nu$,
          },
        display columns/7/.style=
          {
          column name=$-\alpha$,
          },
        display columns/8/.style=
          {
          column name=$\sigma_\alpha$,
          },
        display columns/9/.style=
          {
          column name=$\chi^2_\nu$,
          },
        display columns/10/.style=
          {
          column name=$-\alpha$,
          },
        display columns/11/.style=
          {
          column name=$\sigma_\alpha$,
          },
        display columns/12/.style=
          {
          column name=$\chi^2_\nu$,
          },
        display columns/13/.style=
          {
          column name=$-\alpha$,
          },
        display columns/14/.style=
          {
          column name=$\sigma_\alpha$,
          },
        display columns/15/.style=
          {
          column name=$\chi^2_\nu$,
          },
        fixed,
        fixed zerofill,
        precision=2,
        string replace={0.00}{},
        string replace={0}{},
        column type=r,
        every nth row={7}{before row=\midrule}
    ]{tables/csv/82763272.csv} % filename/path to file
  \end{center}
\end{table*}

\clearpage

\section{Fourth quasar details, J023259.60+004801.7   (Thing ID = 110619413)} \label{sec:appendixC}

\begin{figure*}
    \centering
    \includegraphics[width=\textwidth]{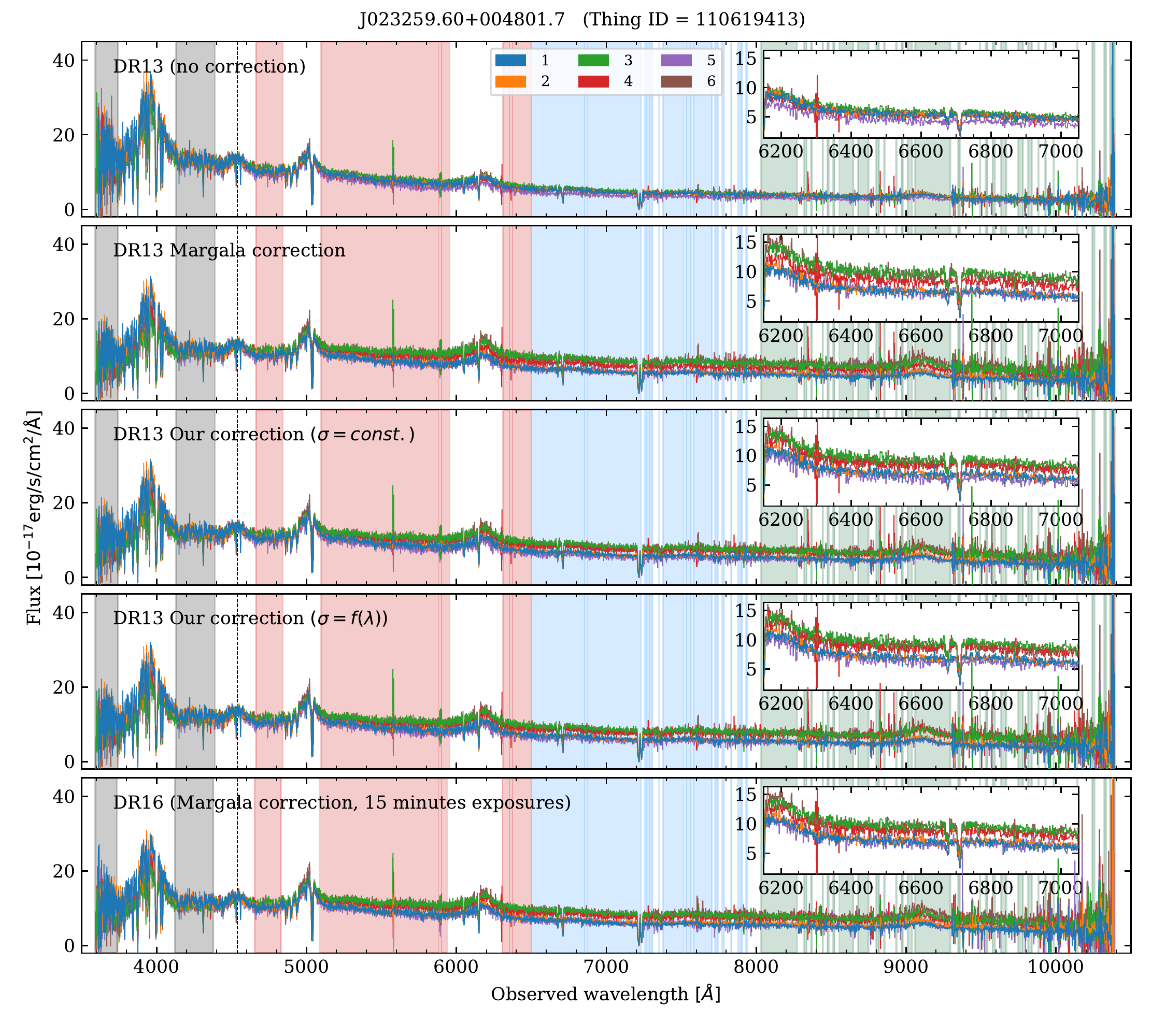}
    \caption{Same as \figref{fig:spec_obs} except for the quasar J023259.60+004801.7.}
    \label{fig:J023259correctionsB}
\end{figure*}

\begin{figure*}
    \centering
    \includegraphics[width=\textwidth]{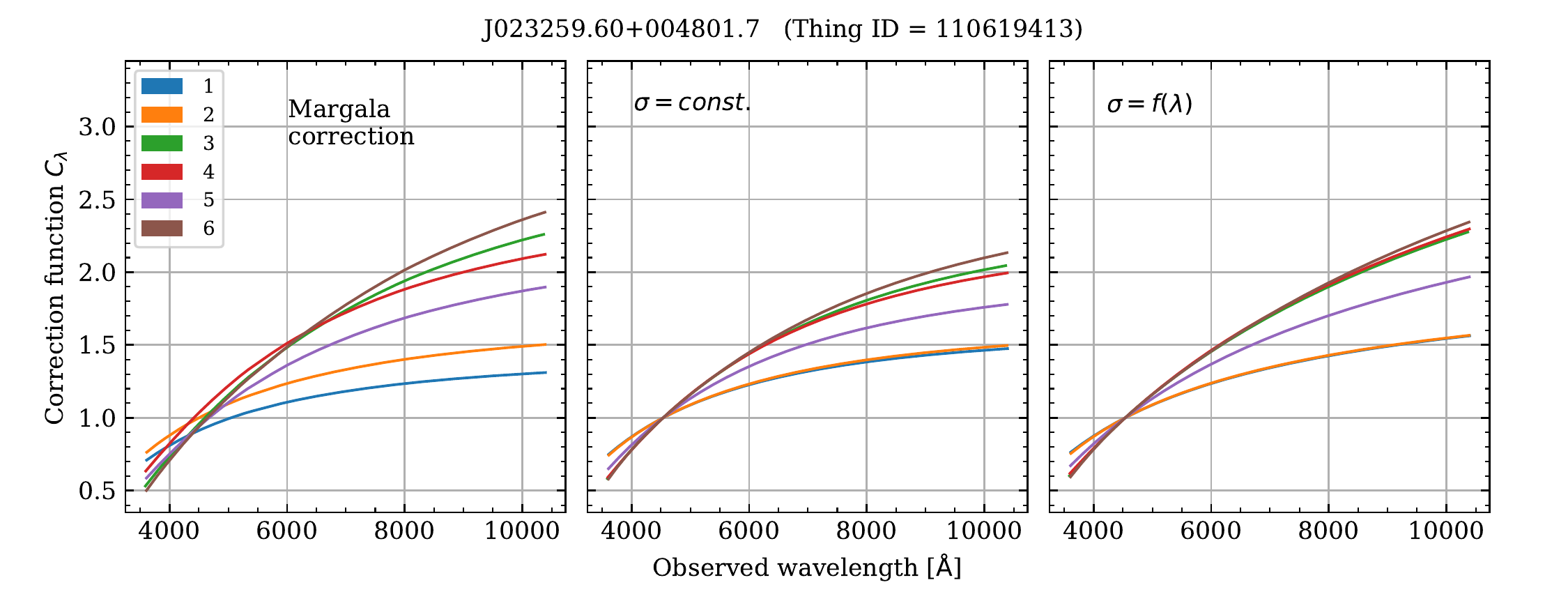}
    \caption{Same as \figref{fig:corr} except for the quasar J023259.60+004801.7.}
    \label{fig:110814178_calib_corr_comparison}
\end{figure*}

\begin{figure*}
    \centering
    \includegraphics[width=\textwidth]{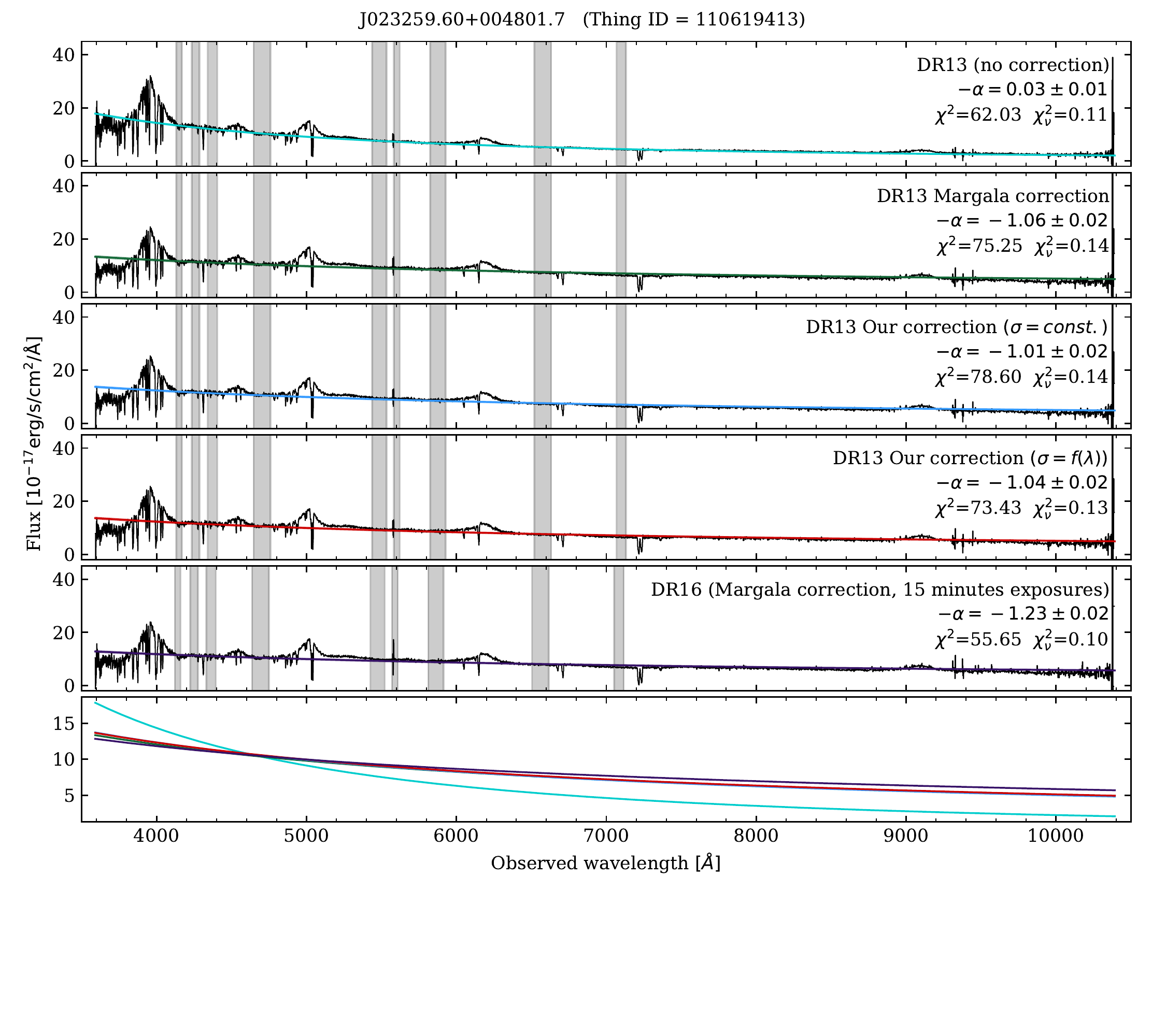}
    \caption{Same as \figref{fig:97176486_mean_spectrum_comparison} except for the quasar J023259.60+004801.7.}
    \label{fig:J023259corrections1}
\end{figure*}

\begin{table*}
  \begin{center}
    \caption{Same as Table \ref{tab:consistency_97176486} except for the quasar J023259.60+004801.7.
%Comparing agreement between individual exposures for the SDSS quasar J023259.60+004801.7 (Thing ID 110814178) for each correction. The first two columns give the wavelength ranges over which the comparison statistic, $\xi^2$ (Eq.\eqref{eq:xisquared}) are made. The subsequent columns give $\xi^2$ and its estimated uncertainty, $\sigma(\xi^2)$, calculated from the number of degrees of freedom, Eq.\eqref{eq:sigxisq}.
}
    \label{tab:consistency_110814178}
    \begin{filecontents*}{tables/csv/110814178_variances.csv}
   3500.00,   4538.00,   2652.68,     29.60,   3359.50,     29.60,   3032.71,     29.60,   3007.58,     29.60,   3163.16,     29.26
   4538.00,   6500.00,   8303.36,     43.47,  19114.31,     43.47,  15030.70,     43.47,  15271.90,     43.47,  17509.87,     43.50
   6500.00,   8000.00,   8602.09,     37.20,  32627.67,     37.20,  22317.98,     37.20,  24139.20,     37.20,  24281.14,     37.20
   8000.00,  10400.00,   4436.60,     31.50,  23238.58,     31.50,  14073.37,     31.50,  16989.08,     31.50,  16344.11,     31.50
   3500.00,  10400.00,  23994.72,     71.96,  78340.06,     71.96,  54454.76,     71.96,  59407.77,     71.96,  61298.27,     71.83
\end{filecontents*}
\pgfplotstabletypeset[
      multicolumn names, % allows to have multicolumn names
      col sep=comma, % the separator in our .csv file
      every head row/.style=
        {
        before row={\toprule  \toprule
                     \multicolumn{2}{c}{Wavelength range} & 
                     \multicolumn{2}{c}{Uncorrected} & \multicolumn{2}{c}{Margala} & 
                     \multicolumn{2}{c}{Our correction} & \multicolumn{2}{c}{Our correction} & 
                     \multicolumn{2}{c}{DR16} \\
                     & & \multicolumn{2}{c}{} 
                     & \multicolumn{2}{c}{($\sigma=const.$)} & \multicolumn{2}{c}{($\sigma=const.$)} & \multicolumn{2}{c}{($\sigma=f(\lambda)$)} &
                     \multicolumn{2}{c}{($\sigma=const.$)} \\
                    },
        after row = {\midrule}
        }, 
        every last row/.style={after row=\bottomrule\bottomrule},
        display columns/0/.style=
          {
          column name = $\lambda_{min}$,
          precision=0,
          },
        display columns/1/.style=
          {
          column name=$\lambda_{max}$,
          precision=0,
          },
        display columns/2/.style=
          {
          column name=$\xi^2$,
          },
        display columns/3/.style=
          {
          column name=$\sigma(\xi^2)$,
          },
        display columns/4/.style=
          {
          column name=$\xi^2$,
          },
        display columns/5/.style=
          {
          column name=$\sigma(\xi^2)$,
          },
        display columns/6/.style=
          {
          column name=$\xi^2$,
          },
        display columns/7/.style=
          {
          column name=$\sigma(\xi^2)$,
          },
        display columns/8/.style=
          {
          column name=$\xi^2$,
          },
        display columns/9/.style=
          {
          column name=$\sigma(\xi^2)$,
          },
        display columns/10/.style=
          {
          column name=$\xi^2$,
          },
        display columns/11/.style=
          {
          column name=$\sigma(\xi^2)$,
          },
        fixed,
        fixed zerofill,
        precision=2,
        string replace={0.00}{},
        string replace={0}{},
        column type=r,
        set thousands separator = {\,},
        every nth row={4}{before row=\midrule}
    ]{tables/csv/110814178_variances.csv} % filename/path to file
  \end{center}
\end{table*}

\begin{table*}
  \begin{center}
    \caption{Same as Table \ref{tab:si_97176486} except for the quasar J022954.42-005622.5. \mgt{(EZ: Give DR7 value)}}
    \label{tab:si_110814178}
    \begin{filecontents*}{tables/csv/110814178.csv}
1    ,+0.04,0.02,0.98,-0.58,0.02,1.04,-0.65,0.02,1.05,-0.68,0.02,1.03,-0.72,0.02,1.19
2    ,+0.05,0.02,1.10,-0.64,0.02,1.21,-0.65,0.02,1.22,-0.68,0.02,1.19,-0.83,0.02,1.32
3    ,-0.18,0.02,0.99,-1.62,0.02,1.57,-1.43,0.02,1.49,-1.48,0.02,1.37,-1.62,0.02,1.73
4    ,+0.06,0.02,0.97,-1.19,0.02,1.24,-1.18,0.02,1.26,-1.26,0.02,1.15,-1.33,0.02,1.41
5    ,+0.41,0.02,0.93,-0.83,0.02,1.09,-0.64,0.02,1.06,-0.69,0.02,1.02,-0.91,0.02,1.04
6    ,-0.09,0.02,0.97,-1.64,0.02,1.18,-1.38,0.02,1.14,-1.41,0.02,1.10,-1.62,0.02,1.27
$\mu$,+0.05,0.00,0.99,-1.08,0.00,1.22,-0.99,0.00,1.20,-1.03,0.00,1.14,-1.17,0.00,1.33
$\sigma_\mu$,+0.18,0.00,0.00,+0.43,0.00,0.00,+0.35,0.00,0.00,+0.36,0.00,0.00,+0.37,0.00,0.00
$\sigma_\mu/\sqrt{6}$,+0.07,0.00,0.00,+0.18,0.00,0.00,+0.14,0.00,0.00,+0.15,0.00,0.00,+0.15,0.00,0.00
\end{filecontents*}
\pgfplotstabletypeset[
      multicolumn names, % allows to have multicolumn names
      col sep=comma, % the seperator in our .csv file
      every head row/.style=
        {
        before row={\toprule  \toprule
                     \multicolumn{1}{c}{\multirow{2}{*}{ID}} &  %\multicolumn{1}{c}{\multirow{2}{*}{Observation}} & 
                     \multicolumn{3}{c}{Uncorrected} & \multicolumn{3}{c}{Margala} & 
                     \multicolumn{3}{c}{Our correction} & \multicolumn{3}{c}{Our correction} & 
                     \multicolumn{3}{c}{DR16} \\
                     %& & \multicolumn{3}{c}{} 
                     & & & & \multicolumn{3}{c}{($\sigma=const.$)} & \multicolumn{3}{c}{($\sigma=const.$)} & \multicolumn{3}{c}{($\sigma=f(\lambda)$)} &
                     \multicolumn{3}{c}{($\sigma=const.$)} \\
                    },
        after row = {\midrule}
        }, 
        every last row/.style={after row=\bottomrule\bottomrule},
        display columns/0/.style=
          {
          column name = {},
          string type
          },
     %   display columns/1/.style=
      %    {
      %    column name=(Plate-MJD-Fibre),
      %    string type
      %    },
        display columns/1/.style=
          {
          column name=$-\alpha$,
          },
        display columns/2/.style=
          {
          column name=$\sigma_\alpha$,
          },
        display columns/3/.style=
          {
          column name=$\chi^2_\nu$,
          },
        display columns/4/.style=
          {
          column name=$-\alpha$,
          },
        display columns/5/.style=
          {
          column name=$\sigma_\alpha$,
          },
        display columns/6/.style=
          {
          column name=$\chi^2_\nu$,
          },
        display columns/7/.style=
          {
          column name=$-\alpha$,
          },
        display columns/8/.style=
          {
          column name=$\sigma_\alpha$,
          },
        display columns/9/.style=
          {
          column name=$\chi^2_\nu$,
          },
        display columns/10/.style=
          {
          column name=$-\alpha$,
          },
        display columns/11/.style=
          {
          column name=$\sigma_\alpha$,
          },
        display columns/12/.style=
          {
          column name=$\chi^2_\nu$,
          },
        display columns/13/.style=
          {
          column name=$\alpha$,
          },
        display columns/14/.style=
          {
          column name=$\sigma_\alpha$,
          },
        display columns/15/.style=
          {
          column name=$\chi^2_\nu$,
          },
        fixed,
        fixed zerofill,
        precision=2,
        string replace={0.00}{},
        string replace={0}{},
        column type=r,
        every nth row={6}{before row=\midrule}
    ]{tables/csv/110814178.csv} % filename/path to file
  \end{center}
\end{table*}

\end{document}